\documentclass[preprint]{aastex}

\slugcomment{Accepted for publication in ApJ}

\shorttitle{Outflows in Extreme NLS1s}
\shortauthors{Leighly \& Moore}

\begin{document}


\title{HST STIS Ultraviolet Spectral Evidence of Outflow in Extreme
Narrow-line Seyfert 1 Galaxies: II. Modeling and Interpretation\footnote{Based on observations made
with the NASA/ESA Hubble Space Telescope, obtained at the Space
Telescope Science Institute, which is operated by the Association of
Universities for Research in Astronomy, Inc., under NASA contract NAS
5-26555. These observations are associated with proposal
\#7360.}\hphantom{l}{$^,$}\footnote{Based on observations obtained at Cerro Tololo
Inter-American Observatory, a division of the National Optical
Astronomy Observatories, which is operated by the Association of
Universities for Research in Astronomy, Inc. under cooperative
agreement with the National Science Foundation}}


\author{Karen M. Leighly}
\affil{Department of Physics and Astronomy, The University of
Oklahoma, 440 W. Brooks St., Norman, OK 73019}
\email{leighly@nhn.ou.edu}



\begin{abstract}

We present modeling to explore the conditions of the broad-line
emitting gas in two extreme Narrow-line Seyfert 1 galaxies, using the
observational results described in the first paper of this series.
Photoionization modeling using {\it Cloudy} was conducted for the
broad, blueshifted wind lines and the narrow, symmetric,
rest-wavelength-centered disk lines separately.  A broad range of
physical conditions were explored for the wind component, and a figure
of merit was used to quantitatively evaluate the simulation results.
Of the three minima in the figure-of-merit parameter space, we favor
the solution characterized by an X-ray weak continuum, elevated
abundances, a small column density ($\log(N_H)\approx 21.4$),
relatively high ionization parameter ($\log(U)\approx -1.2$--$-0.2$),
a wide range of densities ($\log(n)\approx 7$--$11$), and a covering
fraction of $\sim 0.15$.  The presence of low-ionization emission
lines implies the disk component is optically thick to the continuum,
and the \ion{Si}{3}]/\ion{C}{3}] ratio implies a density of
$10^{10}$--$10^{10.25}\rm\,cm^{-3}$.  A low ionization parameter
($\log(U)=-3$) is inferred for the intermediate-ionization lines,
unless the continuum is ``filtered'' through the wind before
illuminating the intermediate-line emitting gas, in which case
$\log(U)=-2.1$.  The location of the emission regions was inferred
from the photoionization modeling and a simple ``toy''  dynamical model.
A large black hole mass ($1.3 \times 10^8\,\rm M_\odot$) radiating at
11\% of the Eddington luminosity is consistent with the kinematics of
both the disk and wind lines, and an emission radius of $\sim 10^4 \,
\rm R_S$ is inferred for both.  We compare these results with previous 
work and discuss implications.

\end{abstract}




\section{Introduction}

In 1992, it was demonstrated by Boroson \& Green that the optical
emission line properties in the region of the spectrum around H$\beta$
are strongly correlated with one another.  A principal components
analysis allowed the largest differences among optical emission line
properties to be gathered together in a construct commonly known 
as ``Eigenvector 1''.  The strongest differences hinged on the
strength of the \ion{Fe}{2} and [\ion{O}{3}] emission, and the
width and asymmetry of H$\beta$.  This strong set of correlations is 
remarkable, as it involves correlations among the dynamics and gas
properties between regions separated by vast distances.
Furthermore, the same set of correlations are observed in samples
selected in many different ways.  This pervasiveness leads us to
believe that we are observing the manifestation of a primary physical
parameter. The most favored explanation is that it is the accretion
rate relative to the black hole mass onto the active nucleus.

Narrow-line Seyfert 1 galaxies are identified by their optical
emission line properties.  They typically have narrow permitted
optical lines (FWHM of H$\beta<2000\rm\, km\,s^{-1}$), weak forbidden
lines ([\ion{O}{3}]/H$\beta<3$; this distinguishes them from Seyfert 2
galaxies), and frequently they show strong \ion{Fe}{2} emission
(Osterbrock \& Pogge 1985; Goodrich 1989).  These are the same
properties that define the Boroson \& Green (1992) Eigenvector 1, and
indeed, NLS1s fall at one end of Eigenvector 1.  Thus, the study of
NLS1s offers an attractive research opportunity: if we can understand
the origin of the emission-line properties of NLS1s, then we may be in
a position to  understand AGN emission in general.  Furthermore,
because as least  some of the lines in NLS1s are narrow,
identification of the  frequently strongly-blended lines is less
ambiguous than it is in  broad-line objects. 

This paper is the second in a series of two that explores the UV
emission-line properties of NLS1s.  In the first paper (Leighly \&
Moore 2004; hereafter referred to as Paper I) we introduce the topic
by discussing previous work on the UV emission-line properties of
NLS1s.  We then  present a detailed
analysis of the {\it HST} STIS spectra of two NLS1s, IRAS~13224$-$3809
and 1H~0707$-$495, known for their extreme X-ray properties.  
We found that their continua are as blue as that of the
average quasar.  We observed that the high-ionization emission lines
(including \ion{N}{5} and \ion{C}{4}) are broad (FWHM$\approx\rm
5000\rm \,km\,s^{-1}$), and strongly blueshifted, peaking at around
$2500\rm \,km\,s^{-1}$ and extending up to almost $\sim 10,000\rm \,
km\,s^{-1}$.  In contrast, the intermediate- and low-ionization lines
(e.g., \ion{C}{3}] and \ion{Mg}{2}) are narrow (FWHM
1000--1900$\rm\,km\,s^{-1}$) and centered at the rest wavelength.
\ion{Si}{3}] is prominent, and other low ionization lines (e.g.,
\ion{Fe}{2} and \ion{Si}{2}) are strong.  Based on these observations,
the working model that we adopted considers that the blueshifted
high-ionization lines come from a wind that is moving toward us, with
the receding side blocked by the optically thick accretion disk, and
the intermediate- and low-ionization lines are emitted in the
accretion disk atmosphere or low-velocity base of the wind\footnote{We
do not know with certainty the geometrical and physical origin of the
emission lines in the objects we are discussing here.  However, for
simplicity, we refer to the highly blueshifted high-ionization lines
as originating in the ``wind'', and the narrow, symmetric
low-ionization lines as originating in the ``disk''.}.

The strongly blueshifted \ion{C}{4 } profile suggested that it is
dominated by emission in the wind. Following Baldwin et al.\ (1996), we
used the \ion{C}{4} profile to develop a template for the wind.  We
then used this template, plus a narrow and symmetric component
representing the disk emission, to model the other bright emission
lines. We inferred that the high-ionization lines \ion{N}{5} and
\ion{He}{2} are also dominated by wind emission, and a part of
Ly$\alpha$ is emitted in the wind. A part of Ly$\alpha$ and the
intermediate- and low-ionization lines \ion{Al}{3}, \ion{Si}{3}],
\ion{C}{3}], and \ion{Mg}{2} are dominated by disk emission.  The
1400\AA\/  feature, comprised of \ion{O}{4}] and \ion{Si}{4} was
difficult to model; however, it appears to include both disk and wind
emission.  

IRAS~13224$-$3809 and 1H~0707$-$495 have distinctive X-ray
properties among NLS1s; in order to determine whether their
distinctive properties carry over to the UV, we analyzed {\it HST}
archival spectra from 14 other NLS1s with a range of two orders of
magnitude in UV luminosity.  We find indeed that these two objects are
extreme in this sample in the following properties: strongly
blueshifted \ion{C}{4} line, the low equivalent widths of many of the
lines, particularly \ion{C}{4} and \ion{He}{2}, high
\ion{C}{3}]/\ion{C}{4}, \ion{Si}{3}]/\ion{C}{3}],
\ion{Al}{3}/\ion{C}{3}], and \ion{N}{5}/\ion{C}{4} ratios, steep
$\alpha_{ox}$, and blue UV continuum.  Correlation analysis finds a
number of strong correlations.  An anticorrelation between \ion{C}{4}
asymmetry and equivalent width suggests that the line is generally
composed of a highly asymmetric wind component and a narrower
symmetric component.  The anticorrelation between \ion{C}{4} asymmetry
and $\alpha_{ox}$ and $\alpha_{u}$, and with \ion{He}{2} suggests that
UV-strong and X-ray weak continua may be associated with a wind, as
would be expected if the acceleration mechanism is radiation-line
driving.  The dominance of \ion{Al}{3} and \ion{Si}{3}] over
\ion{C}{3}] suggests possibly that the continuum is transmitted
through the wind before it illuminates the intermediate- and
low-ionization line emitting gas.

In this paper, our goal is to use modeling to explore the physical
conditions of the line-emission gas. In \S 2, we compute the predicted
equivalent widths and ratios of the bright UV emission lines using the
photoionization code {\it Cloudy} (Ferland 2001), and then compare
those with the observed values from IRAS~13224$-$3809 to constrain the
ionization parameter, density, covering fraction, column density,
metallicity, and
shape of the photoionizing continuum.  In \S 3, we assume a black hole
mass, and use the photoionization results to estimate the distance
from the central engine to the line emitting region.  We obtain the
distance directly for the disk emission, but need to construct a toy
dynamical model to obtain the distance for the wind.  In \S 4, we
compare with previous results,  discuss broader implications, and
present a speculative scenario to tie all the results together.

\section{Photoionization Analysis}

In Paper I, we showed that the broad-line region lines in the spectra
from IRAS 13224$-$3809 and 1H 0707$-$495 can be fairly robustly
separated into a broad, blueshifted high-ionization component and a
narrow, symmetric intermediate- and low-ionization component.  Thus,
these spectra present us with a rare opportunity to model these
components separately, and obtain physical constraints for the
line-emitting gas in the two different emission regions.

We use the photoionization code {\it Cloudy} Version 94.00 (Ferland
2001) to simulate the emission lines.  Our approach is to simulate the
line emission in gas with a range of densities, column densities and
metallicities subject to a range of photoionizing fluxes, assuming
several different continuum shapes.  We then compare the predicted
line equivalent widths and ratios with the observed values.  Our
approach is similar to that described by Baldwin et al.\ (1996),
although it differs somewhat in the details of the method used for
comparing the observed and predicted line properties, as described
below.

\subsection{Continuum}

The lines emitted by AGN should depend on the shape of the continuum
(Krolik \& Kallman 1988; Casebeer \& Leighly 2004).  Therefore, the
first step in the photoionization modeling is to construct the
continuum.   Our two objects are essentially identical; therefore, since
there is more multiwavelength information is available for
IRAS~13224$-$3809 than for 1H~0707$-$495, we concentrate on developing
a continuum for IRAS~13224$-$3809 and modelling it in detail.

There was no X-ray observation coordinated with the {\it HST} observation,
so there is uncertainty in extrapolation of the {\it HST} spectra to the
non-simultaneous {\it ROSAT} and {\it ASCA} data.  However, both
IRAS~13224$-$3809 and 1H~0707$-$495 show high amplitude X-ray
variability on time scales of a few thousand seconds (e.g., Leighly
1999a; Boller et al.\ 1997) and defining an average X-ray continuum
would be ambiguous even if simultaneous observations were available.
For IRAS~13324$-$3809, we used the average {\it ASCA} spectrum from
the observation conducted 1994 July 30 and the average {\it ROSAT}
PSPC spectrum from the 1992 August 10 observation.  This produced a
continuum spectrum with $\alpha_{ox}$, defined here in the standard
way as the index of the power law stretching between 2500\AA\ and
2~keV (e.g., Wilkes et al.\ 1994), of 1.63 for IRAS~13224$-$3809.  We
note that for 1H~0707$-$495, a comparison between the {\it HST} spectrum
and the {\it ASCA} spectrum obtained 1995 March 15 reveal also
$\alpha_{ox}=1.63$.  An average quasar with a log luminosity at
2500\AA\ of 29.3 is expected to have $\alpha_{ox}=1.4$ with a range
for most of the objects being between 1.06 and 1.64 (Wilkes et al.\
1994).  Therefore, IRAS~13224$-$3809 and 1H~0707$-$495 appear to be
somewhat X-ray weak.

Much of the continuum is not well observed or unknown and therefore we
adopt the generic ``AGN continuum'' included in {\it Cloudy} for the
energy bands where there is no information (Fig.\ 1).  The shape of
the spectrum between 2 and 10 keV is not very well defined for these
objects because of the complexity of the spectrum and low signal to
noise.  Furthermore, nothing is known above 10 keV.  We assume a power
law \footnote{$P(f) \propto f^{-\alpha}$} with $\alpha = 1$ between 1.9
and 100 keV and a power law with $\alpha = 1.6$ greater than 100 keV.
IRAS~13224$-$3809 has a large excess of infrared emission compared
with the AGN continuum (Fig.\ 1) but as noted in Paper I, this object
seems to have a strong galaxy component and therefore the infrared
emission may be dominated by the starburst.  Therefore, we adopt the
{\it Cloudy} AGN continuum longward of  4430 \AA\/.

\begin{figure}
\epsscale{1.0}
\plotone{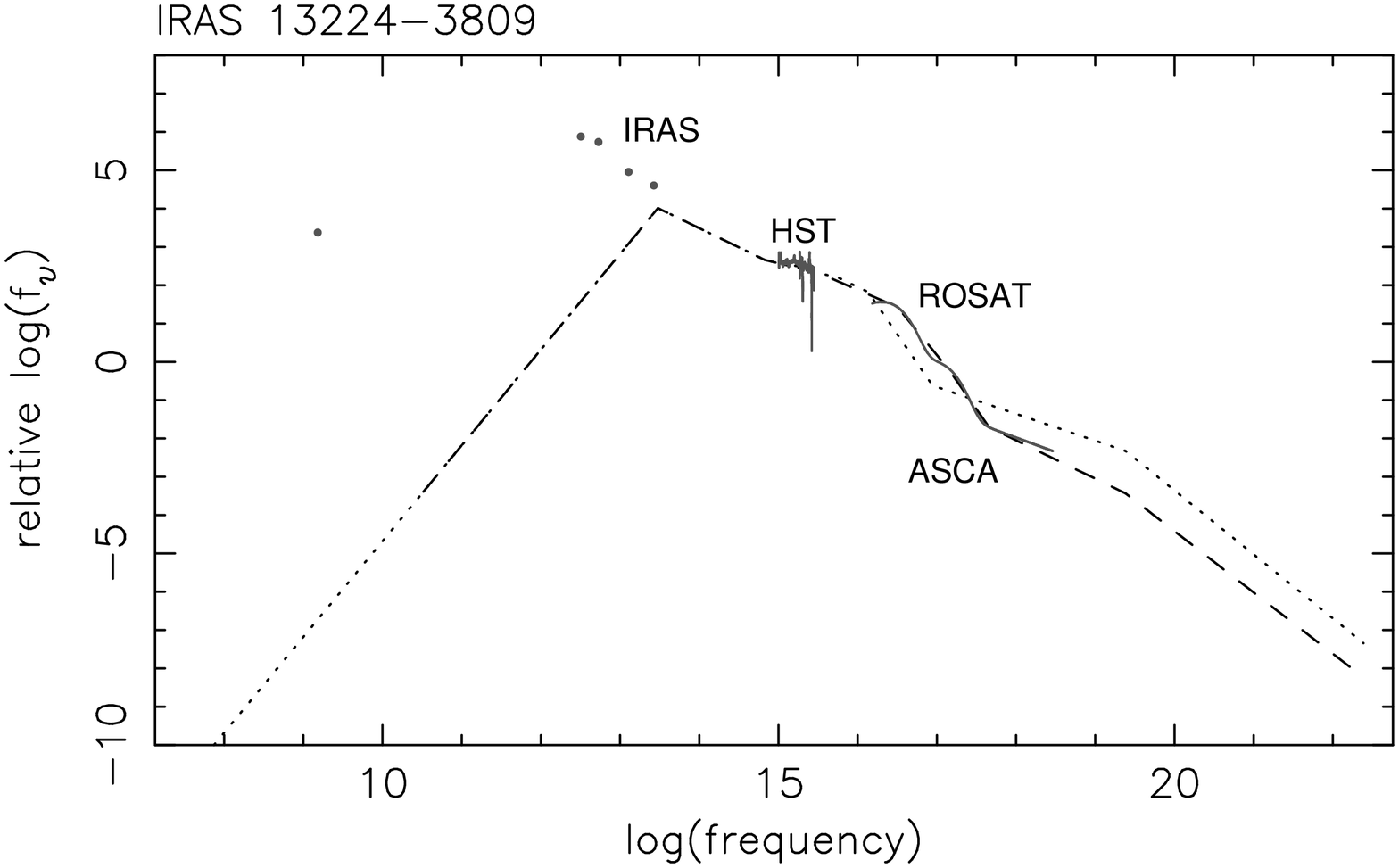}
\caption{The adopted continuum for {\it Cloudy} simulations (dashed
  line) was constructed from the nonsimultaneous {\it HST}, {\it ASCA}
and {\it ROSAT} observations, and the {\it Cloudy} ``AGN'' continuum,
shown by the dotteded line.\label{fig1}}
\end{figure}

\placefigure{fig1}

The {\it ROSAT} soft X-ray spectrum was rescaled to match the {\it
ASCA} flux level, and power laws with slopes $\alpha$ of 2.5 and 2.96
were used between 0.13--0.58 keV and 0.58--1.9 keV, respectively.
Between 0.011 and 0.13 keV, the spectrum is unknown. Rather than try
to guess a spectral shape in this region, we initially simply joined
the UV and X-ray spectra, using a power law with slope 0.9.   The resulting
continuum is   shown in Fig.\ 1, compared with the {\it Cloudy} AGN
continuum.  It is not surprising that the adopted continuum is
brighter in soft X-rays but fainter in hard X-rays than the {\it
Cloudy} AGN continuum.  

We also considered three other related continua.  For two of the
continua, we consider the possibility that the X-ray flux is lower
than it was during the {\it ASCA} observation.  This may be justified
by the observed high amplitude variability of these objects, and the
fact that 1H~0707$-$495 has been observed in a low state, about a
factor of 10 lower than during the {\it ASCA} observations, in three
out of the six epochs of observation since 1990 (Leighly et al.\ 2002;
Leighly et al.\ in prep.). Furthermore, it is quite possible that, if
the line-emitting gas has a disk-like configuration and lies close to
the accretion disk, it may see a continuum deficient in X-rays.  We
constructed the X-ray weak continua by decreasing the X-ray flux by an
arbitrarily-chosen factor of 4.  Since we do not know where the break
to the X-ray spectrum occurs in the unobservable EUV, we tried two
low-flux continua: one, called ``low flux (soft)'' continuum, in which
the {\it HST} spectrum is directly joined by a power law to the X-ray
spectrum, and another, called ``low flux (hard)'' continuum, in which
the slope of the {\it HST} spectrum is extrapolated to 77~eV, the
ionization potential to create $N^{+4}$, before breaking to the
X-rays.  These continua are shown compared with the nominal
IRAS~13224$-$3809 continuum in Fig.\ 2.

Finally, it has been proposed that gas responsible for absorption
lines in Broad Absorption Line QSOs can be accelerated by radiative
line driving if the X-rays are absorbed out of the continuum before it
illuminates the absorbing gas (Murray et al.\ 1995).  We therefore
also try the IRAS~13224$-$3809 continuum transmitted through gas with
$\xi=200$\footnote{$\xi$ is defined by $\xi=L/nR^2$, where L is the
luminosity between 1 and 1000 Ryd, $n$ is the gas density, and $R$ is
the radius (Tarter, Tucker \& Salpeter 1969).} (which corresponds to
$\log(U)=0.9$) and column density of $10^{23}\rm \, cm^{-2}$.  The
ionization parameter of $\xi=200$ was chosen because it is close to
the largest ionization parameter possible with negligible loss of
radiation-driving force (Stevens \& Kallman 1990).  The continuum is
shown in Fig.\ 2.  Note that when discussing transmitted continua, we
report the photon fluxes and ionization parameters {\it before}
transmission. 

\subsection{Modeling the Wind lines}

We next use {\it Cloudy} to compute the emission line spectrum of gas
subject to this continuum.  We concentrate on the following ten
emission lines: Ly$\alpha$, \ion{N}{5}, \ion{Si}{4}, \ion{O}{4}],
\ion{C}{4}, \ion{He}{2}, \ion{Al}{3}, \ion{Si}{3}],
\ion{C}{3}], and \ion{Mg}{2}. 

The intent of the photoionization modeling described in this and the
next section is to use the emission line ratios and equivalent widths
to obtain estimates of the ionization parameter, column density,
metallicity, density, and covering fraction, and constraints on the
illuminating continuum as is possible.  These results are then used in
the subsequent section to constrain the remaining parameters as well as
the distance of the line-emitting gas from the continuum source. 

We first investigate the broad, blueshifted ``wind'' lines.  Of the 10
lines listed above, broad components of the first 6 (Ly$\alpha$,
\ion{N}{5}, \ion{Si}{4}, \ion{O}{4}], \ion{C}{4}, \ion{He}{2}) were
inferred to be present (measurements given in Table 2 of Paper I), and
the last 4 were not detected.  We prescribe the upper limit of 5 \AA\
to \ion{Al}{3}, \ion{Si}{3}], and \ion{C}{3}], and an upper limit of
10 \AA\ to \ion{Mg}{2}.  Generous upper limits were used because a
weak broad component could not be easily distinguished from the
continuum.

There are six parameters that influence the predicted equivalent
widths of the emission lines: the ionization parameter, the gas
density, the column density, the covering fraction, the metallicity,
and the continuum.  We have already described the four continua that
we consider.  We consider three metallicity combinations: solar; all
metals enhanced by a factor of 5 over solar; and all metals enhanced
by a factor of 5 over solar, with nitrogen alone enhanced by a factor
of 10 over solar.  We do not make a complete search of metallicity
parameter space because preliminary investigations show that the data
cannot constrain the metallicity with high precision; this issue will
be revisited when the {\it HST} spectra of 1H~0707$-$495 is analyzed
in combination with the {\it FUSE} spectra (Leighly et al.\ in
prep.).  The combinations to present were determined on the basis of
preliminary trial-and-error batches of simulations.  
We make grids of the equivalent widths of the ten emission lines
listed above predicted from {\it Cloudy} runs for the following three
parameters, followed by the range and interval that we consider:
$\log$(density) [7.0-12.0 by 0.25]; $\log$(U) [ionization 
parameter\footnote{The ionization parameter $U$ is defined as the
photoionizing flux $\Phi$, divided by the density $n$ and $c$, the
speed of light, to make it dimensionless.  It can be thought of as the
ratio of the photon density, controlling the rate of ionization, and
the matter density, controlling recombination.}, $-3.0$--0 by 0.25]; 
$\log(N_H^{max})$ [21--23.4 by 0.2], where $N_H^{max}$ is defined as
$\log(N_H)-\log(U)$, and $N_H$ is the column density.  $N_H^{max}$
is a useful parameter for studies of photoionized gas, because it
adjusts the column density to account for the ionization parameter.
Thus, for a particular value of $N_H^{max}$, we probe to the same
depth in terms of elemental ionization fractions regardless of the
ionization parameter.  The range of ionization parameter explored
encompasses the ``nominal'' value understood for the AGN broad-line
region of $-2$, as well as lower values inferred for NLS1s by
Kuraszkiewicz et al.\ (2000).  The range of $\log(N_H^{max})$
considered spans gas that is rather optically thin in the continuum
through gas that includes the hydrogen ionization front (around
$\log(N_H^{max})$ of 23).  The range of densities considered starts
just above the constraint imposed by the lack of observed broad
component of [\ion{O}{3}] ($n_{crit}\approx 10^6\rm \, cm^{-3}$;
Osterbrock 1989) through densities at which the bright permitted lines
start to become thermalized (e.g., Kuraszkiewicz et al.\ 2000).   In
computing the equivalent width, we explore a range of covering
fractions [0.01--0.5   by 0.01].   

Initially, our search through these thousands of simulations was made
by an iterative process of running simulations, comparing results
with the data, understanding why particular solutions didn't work,
tweaking the parameters, and running more simulations.  Finally,
however, we devised a less subjective criterion to determine how well
the parameters fit the data.  We use a ``figure of merit'', hereafter
abbreviated {\it fom}, that is defined as sum of the absolute values
of the differences of the log of the equivalent width of the emission
lines observed, and the log of the emission line equivalent widths
predicted by {\it Cloudy}.  The lines considered in the sum were the
ones detected in the wind: Ly$\alpha$, \ion{N}{5}, \ion{O}{4},
\ion{C}{4}, and \ion{He}{2}.  We then exclude any solutions from
consideration if they exceed the upper limits for the lines not
detected in the wind: \ion{Al}{3}, \ion{Si}{3}], \ion{C}{3}, and
\ion{Mg}{2}.  Clearly, the {\it fom} will be small when the predicted
equivalent widths are close to the observed values. In this way, it is
like a surrogate $\chi^2$; however, {\it fom} is not a proper 
statistical parameter like $\chi^2$. Thus, we computed the {\it fom}
for each simulations, constructing a four-dimensional {\it fom} matrix
for each of the  twelve combinations of continuum and metallicity.
We then searched the {\it fom} matrices for the minimum values of {\it
fom}.   

There are two ways to think about the combination of parameters
producing low values of {\it fom}.   A first way  is to look for the
solution that produces 
the absolute smallest value of {\it fom}.  For every combination of
continuum and metallicity, the smallest {\it fom} is produced in very
high density ($\log(n) \approx 12$), very high column
($\log(N_H^{max})\approx 23.4$),
very high ionization ($\log(U)\approx 0$) gas.  This is essentially the
solution that Kuraszkiewicz et al.\ 2000 found, and we discuss the
nature of this solution as Type I solutions below.

A second way to view the solutions takes a different view of the
physical conditions in the emission region.  In reality, the line
emitting gas may not be characterized by a {\it single} value of the
parameters; in reality, a {\it range} of densities and ionization
parameters may be present.   In this case, we may be more interested
in {\it fom} minima that are shallow, implying that combinations of
parameters neighboring the {\it fom} minimum also should produce
emission lines with equivalent widths and ratios near the observed
values.  A solution characterized by a broad {\it fom} minimum relies
less on fine-tuning than does one with a narrow minimum.  

To first explore the {\it fom} minima for the twelve combinations
of metallicity and continuum, we list in Table 1 the number of points
in the {\it fom} grid that fall below an arbitrarily chosen value of
120, as well as the range of parameters encompassed by {\it fom}$<120$.
We find that a much larger region of parameter space produces emission
lines and ratios close to the observed values when the metallicity is
elevated, and the X-ray flux is low.  

Why are more solutions available for the higher metallicity and lower
X-ray flux continua?  The reason seems to be the presence of multiple
minima in the high-metallicity, low-X-ray-flux {\it fom} matrices.  We
identified these multiple minima in two ways.  We looked at the
distribution of parameters for solutions where  {\it fom}$<120$.  Double
peaks in several cases indicated multiple minima.  We also looked at
histograms of the ``distance'' of the solutions from the solution
characterized by global minimum {\it fom}, where the distance was simply
defined by the square root of the sum of the squares of the parameter
indices.  Multiple peaks were found in these histograms, again
indicating the presence of multiple minima.  We determined the
density, ionization parameter, column density, and covering fraction
responsible for the different minima by looking at histograms  of
parameters selected from small ranges  in the {\it fom} distances, and
we could identify the local minimum by looking at the {\it fom}
rank-ordered list of parameters.  

We found that the minima could be divided into three different types,
based on the physical conditions of the gas.  To understand the
properties of these minima, two plots are presented.  Fig.\ 3 shows
the log of the ratio of the observed equivalent widths to the
predicted equivalent widths for cuts through the equivalent width
matrix that intersect representative examples of the three types of
minima.  Fig.\ 4 shows contours of {\it fom} up to the value of $120$
for planes of $\log(n)$ and $\log(U)$, and $\log(N_H^{max})$ and
covering fraction for the same representative minima.

\begin{itemize}
\item {\bf Type I: High density, high $N_H^{max}$} This solution is
  similar to the ones obtained without the benefit of profile
  deconvolution by Kuraszkiewicz et al.\ 2000.  The representative
  case shown in Fig.\ 3 and Fig.\ 4, has the nominal continuum and
  solar metallicity, and $\log(n)=12$, $\log(U)=0$,
  $\log(N_H^{max})=23.4$, and the covering fraction is 0.26.  The
  solutions are pegged at the highest values of density, ionization
  parameter, and column density.  As can be seen in Fig.\ 3, the very
  high density results in some degree of thermalization in such
  permitted lines as Ly$\alpha$ and \ion{C}{4}, reducing the
  equivalent widths of these lines.  The high column ensures
  sufficient line emission, especially from lower-ionization
  species, such as \ion{Si}{4}.  The high ionization decreases
  emission from lower-ionization lines.

  This solution is rather fine-tuned.  It encompasses only a small
  region of parameter space constrained at high densities, as seen in
  Fig.\ 4, because lower density gas produces too much much Ly$\alpha$
  and \ion{C}{4}.  Lower ionization parameter gas produces too much
  \ion{C}{3}].

  As discussed above, this type of solution always produces the lowest
  {\it fom} for all combinations of continuum and metallicity.  The
  parameters corresponding to this minimum are nearly the same for the
  other continua and nominal metallicity.  As metallicity is enhanced,
  the range of ionization parameter allowed decreases because
  intermediate-ionization lines are stronger; however, the column
  density constraint is looser because the hydrogen ionization front
  moves toward lower column densities as the metallicity increases.

\item {\bf Type II: Intermediate density, intermediate column density,
  intermediate ionization parameter}.  This solution is different
  from Type I in   that the density is high enough for only Ly$\alpha$
  to start to be   attenuated by thermalization.   The ionization
  parameter is high   enough, and the column density is   low enough that the
  intermediate-ionization line upper-limit constraint is met, and
  parameter space is constrained sharply toward lower densities and
  higher column densities by the upper limit on \ion{C}{3}] emission
  in the wind.  The representative solution shown in Fig.\ 3 and
  Fig.\ 4 has the nominal continuum, and metals$\times 5$ and
  nitrogen$\times 10$ metalicity, and $\log(n)=10.75$, $\log(U)=-0.8$,
  $\log(N_H^{max})=22.6$, and covering fraction of 0.08.  

  This solution is found for the nominal, soft low-flux
  continua, and the $\xi=100$ shielded continua when the metals are
  enhanced by a factor of 5 over solar,   and nitrogen is enhanced by
  a factor of 10 over solar.   

\item {\bf Type III: Large range of densities, low column}  This
  solution is different than Type II in that a large range of
  densities, from $\log(n)=7$ to $\log(n)=\sim 11$ are accessible.
  This is because the relatively low column density truncates the
  plasma so that no intermediate-ionization lines are produced.
  Because the column does the job of excluding low-ionization lines,
  a relatively large range of ionization parameters are also permitted
  (from $\log(U)=-1.0$ to $\log(U)=0.0$).  The representative solution
  shown in Fig.\ 3 and Fig.\ 4 has the hard low-flux continuum,
  and metals$\times 5$ and nitrogen$\times 10$ metallicity, and
  $\log(n)=10.25$, $\log(U)=-0.6$, $\log(N_H^{max})=22.2$, and
  covering fraction of 0.15.  
  
  It should be noted that for this solution, \ion{Si}{4} will not be
  strong enough to contribute to the wind.  This
  is not a problem because, as discussed in Paper I, the 1400\AA\
  feature is composed of both   \ion{Si}{4} and \ion{O}{4}], and a
  robust deblending of these lines was not possible. Therefore, we
  interpret this result to imply that in the  context of this solution,
  the wind contributes only \ion{O}{4}]   to the 1400\AA\ feature.

  This solution is found for both the hard and soft low X-ray
  flux continua for the case where metals are enhanced by a factor of
  5 over solar, and nitrogen is enhanced by a factor of 10 over solar.
  Because of the large range of densities and ionization parameters
  accessible, a very large number of parameter combinations produce an
  {\it fom}$<120$.  

\end{itemize}

We find that the results using the ``shielded'' continuum, which has
been transmitted through the $\xi=100$ gas before illuminating the
line-emitting region, are almost indistinguishable from the results
using the nominal continuum (Table 1).  This is because the $\xi=100$
gas removes photons in the soft X-rays primarily.  The soft
X-ray photons are not required to create the ions that emit the lines
we are considering, nor does their loss represent a substantial loss
of energy to the photoionized plasma.   We note that the result may
have been different had we been considering \ion{O}{6} as well, since
O+5 has an ionization potential of $113\rm\, eV$; this will be
considered in a future paper.  Also, further discussion on the role of
shielding is presented in \S 4.3.3.

Why do the different continua (nominal versus low X-ray flux) and
metallicities produce such different results?  We examine this point
through an example shown in Fig.\ 5.  The top panel of Fig.\ 5
shows the equivalent width contours as a function of density and
photon flux for a $N_H^{max}$ of 22.4 and a covering fraction of 0.1,
for the case of nominal continuum and solar metallicity.  These plots
are similar to those previously discussed in the context of LOC
(locally optimally-emitting cloud) models (Baldwin et al.\ 1995;
Korista et al.\ 1997).  Diagonals on these plots lie along a constant 
ionization parameter, and thus the fact that the equivalent width
contours lie along diagonals highlights the primary role of the
ionization parameter in determining line emission. 

As discussed in Paper I, our two objects are characterized by
exceptionally low \ion{C}{4} equivalent widths, and correspondingly
low \ion{C}{4} to Ly$\alpha$ ratios (0.7).  The top panel of Fig.\ 5 shows that
\ion{C}{4} is produced optimally for $\log(U)=-1.7$.  The \ion{C}{4}
to Ly$\alpha$ ratio is 1.2 for this $\log(U)$, much higher than the
observed value.  Increasing the column density would increase
Ly$\alpha$; however, that would violate the \ion{C}{3}] upper limit
for the wind.  Instead, since Ly$\alpha$ changes less with $\log(U)$
than does \ion{C}{4}, we can obtain a lower \ion{C}{4} to Ly$\alpha$
ratio by moving off the peak of \ion{C}{4} optimal emission, toward
either a lower ionization parameter or a higher ionization parameter.
The fact that the very high ionization line, \ion{N}{5} is observed in
the wind means that we should move toward higher ionization parameter.
However, moving toward higher $\log(U)$ to match the
\ion{C}{4}/Ly$\alpha$ ratio introduces another problem: we do not have
sufficient \ion{O}{4} and/or \ion{Si}{4} to explain the 1400\AA\
feature, because the equivalent widths of these lines decrease more
rapidly toward higher ionization than does \ion{C}{4} (Fig.\ 5).

How does changing the metallicity help?  This point is investigated in
the second panel of Fig.\ 5, which shows the ratio of the emission
lines from a model comprised of the nominal continuum, with the metals
enhanced by a factor of 5, to the emission lines produced with the
nominal continuum and solar metallicity.  The column densities for both
simulations are chosen to truncate at the same continuum optical depth.
We find that modifying the metallicity changes Ly$\alpha$ and
\ion{He}{2} little, as indeed it should not.  The other emission lines
are enhanced toward higher photon fluxes; \ion{O}{4} is especially
enhanced.  The reason for this is that the larger metallicity
increases the number of ions per hydrogen ion available to remove the
energy imparted by photoionization.  Thus, the ions responsible for
the emission lines that we are interested in survive at higher fluxes,
rather than transitioning to a higher ionization state, and therefore
emit at higher fluxes.  This effect may be similar to that explored
previously by Snedden \& Gaskell 1999.

How does changing the shape of the continuum help?  This point is
investigated in the third panel of Fig.\ 5, which shows the ratios of
the emission lines produced by the hard low-flux continuum to
those produced by the nominal continuum, both with solar metallicity.
Changing the continuum alters the flux of Ly$\alpha$ and \ion{He}{2}
little.  The other lines are again enhanced toward higher fluxes.  The
reason for this can be traced to the role of the X-ray continuum as
primarily a a source of energy in the photoionized gas.  The hard
low-flux continuum contributes less energy per ionizing photon to the
gas than does the nominal continuum.  This again allows the ions that
are responsible for the emission lines that we are interested in
survive at higher fluxes.

The effect of changing both the metallicity and the continuum is shown
in the lowest panel of Fig.\ 5.  In this case, we also increased the
nitrogen abundance to 10 times the solar value to better match the
observed \ion{N}{5}.  Selective enhancement of nitrogen is justified
on the basis of secondary enrichment processes that may be important
in AGN (e.g., Hamann et al.\ 2002).  The overall result is that now we
can match the observed \ion{C}{4} to Ly$\alpha$ ratio at a higher
ionization parameter than is optimal for \ion{C}{4} production.  At
the same time, because of the enhanced cooling due to higher
metallicity, and reduced heating by the low-X-ray-flux continuum due
to the smaller amount of energy per ionizing photon, the \ion{O}{4} is 
strongly enhanced, sufficient to explain the 1400\AA\ feature.

We examine this result in complementary way in Fig.\ 6, which shows the
ionization fraction for C+3 and O+3 as a function of depth into the
ionized gas normalized by the depth of the hydrogen ionization
front.  This figure, similar to those presented by  Snedden \& Gaskell
(1999), supports the assertion presented above. When the
metallicity is increased, and when the continuum has lower X-ray flux,
the O+2/O+3 front moves toward the illuminated side of the gas slab
relative to the C+2/C+3 front, indicating that the O+3 ion survives
under higher ionization conditions than it does for the nominal
continuum and metallicity.

To summarize, the analysis presented in this section shows that three
types of combinations of continuum, metallicity, density, ionization
parameter, column density and covering fraction match roughly the
observed high-ionization emission lines from the wind without
overproducing the intermediate- and low-ionization lines not seen in
the wind.  We favor the Type III solution, characterized by a large
range of densities ($\log(n)\sim 7$ to $\sim 11$), moderate range of fairly
high ionization parameters ($\log(U)\sim -1.2$ to $-0.2$), low column
density ($\log(N_H^{max}) \sim 22$ to 22.2, and moderate covering
fraction (0.13--0.25), because it is characterized by a broad minimum
in {\it fom} space, so that a large number of combinations of
parameters produce the emission line equivalent widths and ratios that
we observe.

\subsection{Modelling the Disk Lines}

We next attempt to determine the properties of the gas emitting the
relatively narrow, symmetric intermediate- and low-ionization lines.
Most of the intermediate-ionization lines have profiles consistent
with one another, suggesting that they are all produced in the same
gas.   However, as discussed in Paper I, \ion{Mg}{2} and H$\beta$
lines are measurably narrower than the \ion{C}{3}], \ion{Si}{3}] and
\ion{Al}{3} emission lines,  indicating that they may be produced
predominately at larger radii. 

To investigate properties of the disk component, we consider the 10
lines used to constrain the wind. As discussed in Paper I, we do not
see narrow, symmetric components of the high-ionization lines
\ion{N}{5}, \ion{C}{4} and \ion{He}{2}; therefore we set upper limits
for them.  The upper limits for \ion{N}{5} and \ion{He}{2} are set to
1 \AA\/.  This is a tighter constraint than was imposed for the upper
limits on lines not seen in the wind, but it is justified by the fact
that narrow lines are easier to detect.  We have made the simplifying
assumption that the \ion{C}{4} line is produced solely in the wind.
However, a portion of the \ion{C}{4} profile overlaps that of the disk
line \ion{C}{3}], and this portion could in principle be produced in
the disk.  The equivalent width of that overlapping portion of the
\ion{C}{4} profile is 4.1 \AA\/, so we take this as the upper limit of
\ion{C}{4} produced in the disk component.  We also consider the
low-ionization lines \ion{C}{2}$\lambda 1335$ and \ion{C}{2}$\lambda
2327$, both clearly detected, and \ion{N}{3}$\lambda 1750$, which is
marginally detected at best.  We do not use the prominent \ion{Si}{2}
lines, as it is clear that our spectra suffer ``The \ion{Si}{2}
Disaster'' (Baldwin et al.\ 1996); that is, \ion{Si}{2} appears to be
selectively excited in some way.

We set up the model as follows. We use the results from the previous
section, and assume the Type III solution for the wind, comprised of
the hard low-flux continuum and enhanced metallicity (metals enhanced by a
factor of 5, and nitrogen enhanced by a factor of 10 over solar).  In
the disk component, we observe lines with quite low ionization
including \ion{Mg}{2}, \ion{C}{2} and \ion{Si}{2}.  These
low-ionization species suggest that the column density of the disk
component extends past the hydrogen ionization front.  We therefore
set the column density to be 5 times the depth to the hydrogen
ionization front using $N_H^{max}$ as before.  We also considered
higher column densities and found that the results are not sensitive
to this parameter; this is expected since most of the lower-ionization
lines that we consider are produced in the vicinity of the hydrogen
ionization front.

In the disk component, we find the intermediate-ionization
intercombination lines \ion{C}{3}] and \ion{Si}{3}].  These lines are
very useful to constrain the density of the disk component since they
are produced under roughly the same physical conditions (e.g., Hamann
et al.\ 2002), yet they have different critical densities ($3.4 \times
10^9 \rm \,cm^{-3}$ and $1.04 \times 10^{11}\rm \, cm^{-3}$,
respectively).  Therefore, their ratio provides a density constraint
that is fairly independent of ionization parameter.  Requiring the
ratio of these lines to be consistent with the observed value of 1.4,
and requiring the equivalent widths to match the observed values gives
an estimate of the density, the photon flux, and the covering
fraction.  Density and covering fraction could be constrained rather
tightly: they fell between 10.0 and 10.5, and 0.03 and 0.08,
respectively, in all cases (Table 2).  To illustrate these solutions,
we plot the log of the ratio of the predicted equivalent widths to the
measured value or upper limit as a function of the log of the photon
flux (Fig.\ 7).

\placefigure{fig7}

The primary constraint for the disk lines is the low value of the
ratio of the upper limit of the narrow component of \ion{C}{4} to
\ion{C}{3}] compared with other objects discussed in Paper
I. \ion{C}{4} is generally a very strong line.  The two objects
discussed in detail here are remarkable in that the narrow component
of \ion{C}{4} is weak.  Following similar reasoning as presented in \S
2.2, we find that to produce the low inferred ratio of \ion{C}{4} to
\ion{C}{3}], we must move toward lower ionization parameters than that
characterized by optimal \ion{C}{4} emission.  Thus, avoiding
overproducing \ion{C}{4} places an upper limit on the log of the
photon flux of $\sim 17.5$.  This is illustrated in Fig.\ 7: as
the photon flux increases, the predicted flux of \ion{C}{4} increases
fairly rapidly, and quickly becomes much stronger than we see.

Kuraszkiewicz et al.\ (2000) postulated that the low equivalent width
of \ion{C}{4} in NLS1s in general arises because the densities are so
high that \ion{C}{4} is collisionally deexcited.  We find this to be
unlikely.  As discussed above, the ratio of the semiforbidden lines
\ion{Si}{3}] and \ion{C}{3}] provides an estimate of $\log(n)$
between 10.0 and 10.5.  \ion{C}{4} has critical density of $2.06
\times 10^{15}\rm cm^{-3}$ (Hamann et al.\ 2002), and the critical
densities of \ion{Si}{3}] and \ion{C}{3}] are several orders of
magnitude below that.  At densities high enough to deexcite
\ion{C}{4}, we would not expect to see any \ion{Si}{3}] and
\ion{C}{3}] lines at all.  We discuss this point further in \S 4.4.

The low inferred photon fluxes imply a very large radius for the
emission region.  Is there any way it could be smaller?  The reasoning
presented thus far in this section assumes a one-zone model.  But
\ion{Si}{3}] and \ion{C}{3}] are produced optimally at lower
ionization parameters than is \ion{C}{4} (e.g., Korista et al.\ 1997).
We could imagine, then, a radially-extended emission region in which
\ion{C}{4} is produced at smaller radii than \ion{Si}{3}] and
\ion{C}{3}].  In a disk-like geometry, the area of an annulus ($\Delta
A=2 \pi r \Delta r$) increases for a given $\Delta r$ at larger radii.
So one could imagine that if we integrate over a range of radii, we
will accumulate sufficient \ion{C}{3}], due to the larger area of an
annulus at larger radii, to be able to
extend the emission region to smaller radii without violating the
constraint imposed by the \ion{C}{4} upper limit to \ion{C}{3}]
ratio.  Although a detailed  discussion of the implications of a
radially-extended emission region are beyond the scope of this paper,
and will be discussed in future work, we make a quick estimation of
the viability of this hypothesis here. We simply sum the expected flux
from \ion{C}{4}, \ion{C}{3}] and \ion{Si}{3}], weighting the flux as
though the emission were from a disk (i.e., $2\pi r d r$); perhaps
this is appropriate given the very small covering fraction that we
infer.  We perform this summation for a range of starting radii
(corresponding to a range of maximum photon fluxes $\log(\Phi_{max})$)
to a fixed maximum radius corresponding to $\log(\Phi_{min})=15$.  The
results, shown in Fig.\ 8, reveal that with one exception, discussed
below, regardless of the continuum and metallicity, starting the
integration at $\log(\Phi_{max})$ larger than about 18 will produce an
integrated value of \ion{C}{4} that is too large.  Thus, integration
over an extended region still requires a very large radius for the
emission. 

\placefigure{fig12}

There is another, more viable way to decrease the radius of the
emission region, that also conveniently explains the relative emphasis
on lower-ionization emission lines in the disk component.  Consider
the possibility that the continuum has been
filtered\footnote{Henceforth we differentiate between a ``shielded''
continuum, which is assumed to have been transmitted through highly
ionized gas (e.g., Murray et al.\ 1995), and a ``filtered'' continuum,
which is assumed to have been transmitted through the wind while
ionizing and exciting it before illuminating the disk and producing
the observed intermediate- and low-ionization lines.} through the wind
before it illuminates the disk.  Filtering the continuum through the
wind will remove photons in the helium continuum, making the resulting
transmitted continuum unable to produce or excite high-ionization
ions.  We test this possibility as follows.  We use {\it Cloudy} to
obtain the continuum transmitted through a wind with $\log(U)=-0.8$
and $\log(N_H)=21.4$, and metals and nitrogen enhanced by factors of 5
and 10 over solar, respectively.  We used a wind density of
$\log(n)=9.5$ but note that the shape of the transmitted continuum is
not sensitive to this parameter. The value of ionization parameter of
$\log(U)=-0.8$ was chosen because it is characteristic of a large
region of wind parameter space for the Type III solution (Fig.\ 4).
The transmitted continuum is 
shown in Fig.\ 9.  As expected, this continuum is severely deficient
in photons above the helium edge at $54 \rm\, eV$, and therefore it is
neither able to create highly ionized ions, nor to excite them.  We
then used this continuum to illuminate the intermediate- and
low-ionization line emitting gas, and follow the same analysis of the
{\it Cloudy} simulations as above.  The results are shown in
Fig.\ 7 and 10, and parameters are listed in Table 2.  We note that,
as before, we report the ionizing flux before transmission through the
wind in order to use the parameters for distance estimation in \S 4.1.
The net effect of the filtering is to make the continuum quite a bit
softer than the unabsorbed continuum.  The result is a decrease in the
higher ionization lines and an increase in the lower ionization lines.
Thus, \ion{C}{4} is quite a bit weaker overall, which means that we
find the best correspondence between observed and predicted equivalent
widths at a significantly higher ionizing flux than before
($\log(\Phi) \approx 18.7$, corresponding to $\log(U)=-2.1$).

In both cases, we find the predicted value of the equivalent width of
\ion{Mg}{2} to be about a factor of three higher than observed.  As
discussed in Paper I, the \ion{Mg}{2} line is narrower than the
\ion{C}{3}] and \ion{Si}{3}] lines.  Therefore, it is probably
primarily produced at larger radii.  However, that does not alleviate
the problem that a substantial amount of \ion{Mg}{2} is predicted to
be produced in the same gas that emits \ion{Si}{3}] and \ion{C}{3}],
and that is simply not seen.  Kuraszkiewicz et al.\ (2000) also found
\ion{Mg}{2} to be too strong; this seems to be a generic problem for
\ion{Mg}{2}.  It is possible that the predicted \ion{Mg}{2} is
affected by the approximate treatment of radiative transfer in {\it
Cloudy}, although because Mg+ is a very simple atom, this appears to
be unlikely.  It is possible that the gas producing the
intermediate-ionization lines \ion{Si}{3}] and \ion{C}{3}] is not
optically thick in the continuum and the column density of the gas is
truncated so that little \ion{Mg}{2} is produced in the same region as
\ion{Si}{3}] and \ion{C}{3}].  Another possibility is that dust may be
present at large radii, and it may soak up some of the resonance-line
emission.  Further investigation of this issue is beyond the scope of
this paper.

Ly$\alpha$ is also too strong by a factor of 3--4.  We note that
Ly$\alpha$ is very optically thick, and it may also suffer some
problems with radiative transfer.  In particular, we have already
found evidence that Ly$\alpha$ pumps \ion{Fe}{3} in some objects
(Paper I), and Ly$\alpha$ may also pump \ion{Fe}{2} (e.g., Sigut \&
Pradhan 2003).  These mechanisms serve as sinks of Ly$\alpha$.
Ly$\alpha$ would also be severely effected if dust were present.
Further investigation of this issue is beyond the scope of this paper.

As discussed in Paper, I, the 1400 \AA\/ feature is comprised of
\ion{Si}{4} and \ion{O}{4}], with contributions from both the disk and
from the wind.  Fig.\ 7 shows that the simulated \ion{Si}{4} is nearly
as strong as observed at our best value of the photon flux, but
\ion{O}{4}] is more favorably produced at higher photon fluxes and is
therefore too weak.  Therefore, we infer that the disk part of the
1400\AA\/ feature is predominantly \ion{Si}{4}, and the broad part is
predominately \ion{O}{4}], as discussed in \S 2.2.

\ion{Al}{3} is too weak in both models; however, it is only a factor
of 3 too weak when the continuum is filtered through the wind.  This
is to be expected: \ion{Al}{3} increases as the spectrum becomes
globally softer (Casebeer \& Leighly 2004). A few test runs show that
it could be enhanced further by a differential increase in density
with depth into the slab. The increase in density lowers the
ionization parameter, increasing the lower ionization lines that are
strong in these objects.  Another possibility is that the disk is
heated by some mechanism other than photoionization, which might cause
the hydrogen ionization front to retreat farther back into the slab,
allowing stronger intermediate ionization lines.

The ratio of \ion{N}{3}]$\lambda 1750$ to \ion{C}{3}] is considered to
be a good indicator of the relative nitrogen abundance, because these
two lines are produced under the same conditions (Hamann et al.\
2002).  The measured value of \ion{N}{3}], only marginally identified
in the spectrum of IRAS~13224$-$3809, is roughly consistent with the
predicted value for solar metallicities and for metallicity enhanced
by a factor of 5 at our best value of ionizing flux (Fig.\ 7).
However, when the nitrogen is selectively enhanced to 10 times solar,
the predicted emission from \ion{N}{3} is much too strong, by up to a
factor of nearly 10 (Fig.\ 10).  Thus, we find that while the wind
lines require nitrogen to be selectively enhanced by secondary
enrichment, no such enhancement is required for the disk lines.  This
may imply that \ion{N}{5} in the wind is being selectively excited.
It may be pumped by Ly$\alpha$, a process that would seem to be
favorable in a wind due to the large velocity spread and the resulting
line overlap.  Hamann \& Korista (1996) investigated this situation,
and concluded that it was not important for the average quasar, but
the results presented here perhaps suggest that this issue should be
revisited for windy quasars in particular.  It is also possible that
we have not deconvolved the spectrum near Ly$\alpha$ and \ion{N}{5}
completely accurately.  \ion{O}{5} at 1218\AA\/ appears in the
spectrum of RE~1034+39 (Casebeer \& Leighly 2004); we do not model it
here and therefore we may be overestimating the flux of \ion{N}{5}.
However, {\it Cloudy} simulations show that \ion{O}{5} is likely to be
on the order of $1/7.5$ times as strong as \ion{N}{5}.  Furthermore,
that component is likely to blend with the Ly$\alpha$ rather than
\ion{N}{5}.  This issue will be discussed further in a forthcoming
paper on the {\it FUSE} spectrum of 1H~0707$-$495 (Leighly et al.\ in
prep.)

We observe moderately strong features at least partially attributable
to \ion{Si}{2} $\lambda 1260$, \ion{Si}{2} $\lambda 1305$ and
\ion{C}{2} $\lambda 1335$ (Fig.\ 2 in Paper I).  Such strong
\ion{Si}{2} features have also been observed in the NLS1 prototype
I~Zw~1 (Laor et al.\ 1997b) and other narrow-line quasars (Baldwin et
al.\ 1996).  Recently, Bottorff et al.\ (2000) have shown that these
emission lines are predicted to be enhanced under conditions of
turbulence.  We note that a very large turbulent velocity is probably
not warranted in these objects for two reasons.  First, the
intermediate ionization lines are observed to be rather narrow
(although a distribution in turbulent velocities may mask the
broadening effect (Bottorff 2001, P.\ comm.).  Second, the width of
the lines is thought to be related to the Keplerian velocity, which is
the same order as the free-fall velocity, and represents the reservoir
of gravitational energy in the system.  Then, the question of origin
of the energy for very large turbulent velocities arises, since it
should be equal to or less than the equipartition energy of the
gravitational field.  Therefore, we examine the possibility that the
gas is experiencing turbulence with velocities of $2,000\, \rm km\,
s^{-1}$. The primary effect of the turbulence is to reduce the opacity
to resonance lines that are usually just optically thick, and
therefore it is possible that the same effect would be observed if
there were a mild velocity gradient, as might be expected in a disk,
rather than turbulence.

The results are shown in Fig.\ 10.  Compared with the directly
illuminated disk, we find that \ion{Si}{4}, \ion{Al}{3}, and
\ion{C}{2} increase rather dramatically. This is expected because
these lines are formed deep within the slab, near the hydrogen
ionization front, and turbulence has reduced their opacity the most.
However, \ion{C}{4} is still strong, so we are still constrained to
ionizing fluxes of $\sim 17.5$.  For the wind-filtered continuum, we
find a similar increase in \ion{Si}{4} and \ion{Al}{3}.  However,
now \ion{C}{4} again increases strongly with flux, and therefore, we
are constrained to ionizing fluxes less than about $\sim 17.5$.  We
conclude that turbulence (or a velocity gradient originating in
another type of motion) helps to increase the flux of \ion{Al}{3},
which is underproduced in the static models, but then a large radius
for the emission region is required. 

\placefigure{fig10}

To summarize, we find approximate solutions for the intermediate- and
low-ionization disk lines can be provided by gas that is optically
thick to the continuum, with $\log(n)$ between 10 and 10.25 and
covering fraction between 0.035 and 0.07, assuming that the continuum
and metallicity are the best values obtained from the wind modeling.
The low observed ratio of narrow \ion{C}{4} to \ion{C}{3}] constrains
the photoionizing flux to be less than about $\log{\Phi}\approx 17.5$,
if the gas is illuminated directly by the continuum. That value
increases to $\log{\Phi} \approx 18.7$ if the continuum is filtered
through the wind before it illuminating the disk.  Thus, we infer
ionization parameters $\log(U)$ of $-3$ for direct continuum and $-2.1$
for filtered continuum.  Then, the log of the column densities we
infer are $20.7$ and $21.7$, although we note that because gas is
ionization-bounded, larger column densities are not ruled out, and in
fact may be supported by observation of NaD (Thompson 1991) in
IRAS~13224$-$3809.  \ion{Mg}{2} and Ly$\alpha$ are too strong; we
speculate that part of the reason for this may be incomplete modeling
the radiative transfer.  \ion{Al}{3} is somewhat too weak, although
not seriously for the filtered continuum; this may indicate that the
density of the disk increases with depth, or that there is some
turbulent or other motion in the disk.  \ion{N}{3}] is too strong;
this may indicate that we overestimate the nitrogen abundance in the
wind, possibly indicating selective excitation of the wind line
\ion{N}{5} by Ly$\alpha$.

\section{Rough Estimates of Geometry and Kinematics}

In this section, we use the results of the photoionization modeling
and a simple ``toy'' dynamical model to make very rough estimates of
the size scales and geometry of the wind and the disk components.  Our
photoionization modeling is simplified, and the dynamical model
presented here is very simple as well, and cannot be considered a
complete model of the quasar by any means.   Nevertheless, this
exercise is worthwhile because it allows us to obtain quantitative
results that can be used to direct development of more complex models.

First, we make two estimates of the black hole mass.  The Eddington
luminosity is a fiducial parameter of the system; therefore, for the
first black hole mass estimate, we assume that the quasar is radiating
at the Eddington luminosity.  We use cosmological parameters
$H_0=50\rm\,km\,s^{-1}Mpc^{-1}$ and $q_0=0.5$.  We integrate the
continua from $1\,\rm eV$ through gamma-rays, and find that the
bolometric luminosity is plausibly bracketed between $2.2\times
10^{45} \rm \, erg\, s^{-1}$ and $1.5 \times 10^{45} \rm \, erg\,
s^{-1}$, for the nominal and the soft low-flux continua, respectively.
Setting the bolometric luminosity equal to the Eddington luminosity,
we obtain a black hole mass estimate between 1 and 2 $\times 
10^7$ solar masses.  The Schwarzschild radius for a $1.5 \times
10^{7}$ solar mass black hole is $4.4 \times 10^{12}\rm \, cm$.

The second black hole mass estimate is based on the photoionization
modeling performed in the previous section.  The photoionization
modeling results are possibly the least model dependent for the
\ion{C}{3}] line-producing region because of the good density
constraint provided by the \ion{Si}{3}] to \ion{C}{3}] ratio.  As
discussed in \S 2.3, when we assume that the continuum is transmitted
(filtered) through the wind before it illuminates the
intermediate-ionization line-emitting region, we obtain
$\log(n)=10.25$ and $\log(\Phi)=18.7$, yielding an estimate of the
ionization parameter of $\log(U)=-2.1$.  $\Phi$ by itself yields an
estimate of the distance to the emission region of $5.3 \times
10^{17}\rm cm$.  As discussed in Paper I, the FWHM of the \ion{C}{3}]
line from IRAS~13224$-$3809 is $1835 \rm \, km\, s^{-1}$.  The FWHM velocity
of $1835 \rm \, km\,s^{-1}$ corresponds to $\sigma_v \approx
FWHM/2.35=780 \rm \, km\,s^{-1}$.  If the emission region is
flattened, the velocities may be larger, in which case we can use the
effective velocity dispersion $\sigma_v^2=3/4 v_{FWHM}^2$, or
$\sigma_v = 1590\rm \, km\,s^{-1}$ (e.g., Wandel, Peterson \& Malkan 1999).  This
velocity dispersion yields a black hole estimate of $1.0 \times 10^8
\,\rm M_\odot$.  Then, using the calibration of the photoionization
and reverberation methods proposed by Wandel, Peterson \& Malkan
(1999), we obtain a black hole estimate of $1.3 \times 10^8\rm\,
M_\odot$.  This is the value that we finally use; we discuss other
black hole mass values in the appendix. The Schwarzschild radius for
$M_{BH}=1.3 \times 10^8\rm\,M_\odot$ is $3.8 \times 10^{13}\rm \,cm$,
and we infer the object to be radiating at 9--14\% of Eddington
luminosity.

\subsection{Inferences about the Disk}

In this section, we use the photoionization results and the black hole
mass estimates to infer the radius of the intermediate-ionization line
emitting region in Schwarzschild radius units.  We assume that the
continuum illuminates the intermediate-line emitting region after it
is transmitted through the wind; this seems a most promising way to
produce the strong low-ionization line emission observed in NLS1s.
Then, the photoionizing flux is $\log(\Phi)=18.7$, and the distance to
the emitting region is $5.3 \times 10^{17}\rm\, cm$, which corresponds
to $120,000\rm \, R_S$ and $14,000\rm\, R_S$ for the small and large
black hole masses respectively.  Interestingly, these numbers fall
approximately within the range of radii where the accretion disk is
predicted to break up due to self-gravity, and where at least part of
the line emission may occur (Collin \& Hur\'e 2001).

The small black hole mass may not be favored if the velocities of the
intermediate-ionization line-emitting region are of the order of the
Keplerian velocity, or larger, if the emitting region is flattened.
The Keplerian velocity at $120,000\rm \, R_S$ is $610\rm\, km\,
s^{-1}$.  Using $\sigma_v^2=3/4 v_{FWHM}$, this corresponds to
$v_{FWHM}=700\rm \, km\,s^{-1}$, a factor of 2.6 smaller than the observed width of
\ion{C}{3}].  The larger black hole mass doesn't have this problem,
naturally, since the velocity at the line-emitting region is built
into the estimate.

There are a few possible ways to make the small black hole estimate
more plausible.  For example, it is possible that the line width
arises partially in Keplerian motion, and is also partially in 
turbulence.  However, turbulence is unlikely to account for a very
large portion of the line width, because that would require turbulent
velocities in excess of Keplerian velocities and therefore far from
equipartition with the gravitational energy provided by the black
hole.  If the lines are emitted at the point where the disk is
breaking up, the relevant energy is provided by the self-gravity, but
at the break-up point, that should be just about equal to the black
hole gravity, although the mass of the accretion disk may also be
important (Hur\'e 2002).

It is possible that the gas emitting the intermediate- and
low-ionization lines sees a lower flux than we do.  Laor \& Netzer
(1989) show that the ionizing flux has a nearly $\cos(\theta)$
dependence (their Figure 8) for non-rotating black holes.  This may
be important if the disk-line emission region has a flattened
distribution.  A flattened distribution would be consistent with the
small covering fraction, of order of 5\%, that we derive from the
photoionization modeling.  This small covering fraction would then
account for a factor of $\sin(0.05*\pi/2.0)=0.078$ reduction in the
flux, allowing the emission region to have a factor of 3.6 smaller
radius. 

Another possibility is that the flux is attenuated in some way before
reaching the emission region.  We are already assuming that the
continuum is filtered through the wind before illuminating the
line-emitting gas. Further shielding, or absorption of the soft X-rays
in highly ionized gas, a mechanism suggested by Murray et al.\ (1995)
to prevent overionization of the disk wind, is unlikely to help very
much. This is because the steep photon flux spectrum means that most
of the photoionizing photons are found just past 13.6 eV.  This issue
is discussed further in \S 4.3.3.

To summarize, we infer that if the black hole has a relatively small
mass, the line emitting region must see lower flux than we do in order
to reconcile the ionization parameter of the intermediate-ionization
lines with their velocity width, and to place it securely at a smaller
radius than the disk break-up radius.  This may happen naturally if
the emission region is near the plane of the disk, as might be suspected
from the small covering factor.  On the other hand, if the black hole
mass is larger, as inferred from the black hole mass estimate using
the Wandel, Peterson \& Malkan (1999) calibration of the
photoionization and reverberation techniques, the ionization parameter
and velocity width are automatically consistent, and the predicted
radius is consistent with a location interior to or on the order of
the disk break-up radius ($\sim 14,000\rm R_S$).

\subsection{Inferences about the Wind}

We next consider constraints on the wind lines.  Here, we assume that
the wind is best described by the ``Type III'' photoionization
solution for the reasons discussed in \S 2.2.  Thus, we assume
enhanced metallicity (metals a factor of 5 and nitrogen a factor of 10
over solar), and the hard low-flux continuum.  A set of parameters
representative of the broad {\it fom} minimum illustrated in Fig.\ 4
are $\log(U)=-0.8$, the column density $\log(N_H)=21.4$, and the
covering fraction $\sim 0.15$ (Fig.\ 4).  However, the density for this
solution is only loosely constrained at best.  Without the density, we
cannot estimate the distance of the wind emission from the central
engine.  Therefore, in this section we use a toy dynamical model to
roughly constrain the distance and density.

The toy dynamical model is set up as follows.  We assume that
radiative-line driving is responsible for the acceleration.
Radiative-line driving is discussed by a number of authors, especially
in the context of stellar winds (e.g.\ Castor, Abbott \& Klein 1975,
hereafter CAK).  We note that the scenario presented here is somewhat
similar to that proposed by deKool \& Begelman (1995) for broad
absorption line quasars.  We follow the discussion by Arav \& Li
(1994) and Arav, Li \& Begelman (1994), using our observational
parameters to constrain the model.

We postulate that the wind lines are produced in filaments that are
accelerated by radiation pressure.  Filaments need to be confined even
if they are transient.  Filaments can be confined thermally or
nonthermally.  In active galaxies, the predominate confinement
mechanism is unlikely to be thermal. That is, the filaments are
unlikely to be embedded in a hot gas substrate, for many reasons that
have been discussed previously (e.g., Mathews \& Ferland 1987).  Rees
(1987) showed that a nonthermal confinement mechanism may be
plausible. Specifically, a magnetic field may confine the filaments,
and we adopt this scenario here.  Rees (1987) demonstrates that there
should be sufficient magnetic pressure to confine clouds with
densities of $n=10^{10}\rm\,cm^{-3}$ at distances of $10^{18}\rm cm$ for
objects with quasar luminosity.  Our favored densities are about two
orders of magnitude lower at a slightly smaller distance (see below),
so there should be sufficient magnetic pressure to confine the filaments.

We use a highly simplified model to constrain the density and radial
distance from the kinematics.  We solve the equations in one (radial)
dimension; that is, after the gas leaves the disk, it travels in a
straight line.  We neglect Keplerian motion that the gas will have as
it leaves the disk.  These assumptions prevent us from predicting line
profiles.  Our treatment is arguably complimentary to that of e.g.\
Proga, Stone \& Kallman (2000), since we neglect detailed dynamics but include
constraints from the photoionization modeling explicitly.  An
alternative approach that we plan to investigate in the future is
the one used by Kramer, Cohen \& Owocki (2003), who use the profiles of
emission lines in a hot star to derive the wind parameters.

We use the continuity equation,
\.M$=4\pi\,r^2\,\epsilon\,\rho\,v(r)$, where \.M is the outflow
rate, $r$ is the radius, $\epsilon$ is the volume filling factor,
$\rho$ is the mass density, and $v(r)$ is the velocity.  We assume that
$\rho$ is approximately $m_H\,n$, where $n$ is the number density.

The momentum conservation equation is $v dv/dr = (1/\rho) dp/dr -
GM_{BH}/r^2 + a_{rl}$, where $p$ is the pressure, $M_{BH}$ is the
black hole mass, so $GM_{BH}/r^2$ is the acceleration in the
gravitational field of the black hole, and $a_{rl}$ is the acceleration
due to line driving.  As pointed out by Arav, Li \& Begelman (1994)
(hereafter ALB), acceleration due to the pressure gradient should be
small, and we drop that term.  We retain the gravitational
acceleration term.  ALB consider also acceleration due to continuum
driving, but they show that it is generally several orders of
magnitude weaker than the line driving, so we do not consider
the continuum acceleration term here.

The acceleration due to line driving is conventionally written as
$a_{rl}=(n_e \sigma_t F/\rho c) M_L(U,t)$, where $F$ is the photon
flux (e.g.\ CAK) which depends on the radius according to
$1/r^2$. Again we approximate $\rho/n_e=m_H$. Thus $a_{rl}$ can be
understood to be the acceleration due to electron scattering times the
enhancement due to the greater efficiency of resonance line scattering
that is conventionally known as the ``force multiplier''.  The force
multiplier $M_L(U,t)$ depends on the ionization state of the gas, $U$,
since if the ionization is too low or too high, there will be
insufficient resonance-line absorbing ions to scatter the photons and
absorb their momentum.  It also depends on an effective optical depth
parameter $t=\sigma_T\,\epsilon\,n_e\,v_{th}(dv/dr)^{-1}$, where
$v_{th}$ is the thermal velocity of the gas which is typically
$10\,\rm km\,s^{-1}$ in photoionized gas.  As discussed by Arav \& Li
1994 and ALB, this parameter effectively measures the optical depth
within a Sobolev length $v_{th}/(dv/dr)$.  For gas that has a finite
filling factor $\epsilon < 1$, this value is reduced relative to the
conventional value.  Arav \& Li (1994) give a table of $M_L$ as a
function of $U$ and $t$. Those values are computed for the Mathews \&
Ferland (1974) continuum, which is different than our
continuum. However, those values should be approximately appropriate,
except for the fact that our enhanced abundances should increase the
force multiplier.  Thus, while computation of $M_L(U,t)$ is
straightforward, it is beyond the scope of this paper.

The acceleration equation can be easily integrated.  We assign the
integration constant such that $v=0$ at $r=r_f$.  $r_f$ is referred to
as the ``footpoint'' of the flow by Murray et al.\ (1995), and we
adopt that terminology here.     The result is
$v^2=c^2(r_S/r_f-r_S/r)(M_L(U,t)(L/L_{edd})-1)$.

We assume that once the gas leaves the disk, it continues in a
straight line.  We assume that the angle between the observer (assumed
to be along the symmetry axis) and the streamline is $\Theta$.

We use $M_L(U,t)$ to provide the constraint of the physical parameters
of this model, as follows.  The velocity law rises quickly to a
terminal velocity.  We observe a maximum velocity of about $10,000 \rm
\, km\, s^{-1}$.  However, that represents only the component of the
flow parallel to our line of sight, and therefore the terminal
velocity is assumed to be $10,000\,\rm km\,s^{-1}$ divided by
$\cos(\Theta)$.  Given an input foot point $r_f$, and an Eddington
ratio $L/L_{edd}$, we can solve for the $M_L(U,t)$ necessary to
produce that terminal velocity.  Then we use the best $\log(U)$,
inferred to be $-0.8$ as discussed in \S 2.2, and derive $t$ using
other considerations as discussed below.  We then require
self-consistency: we find the radius at which the value of $M_L(U,t)$
tabulated by Arav \& Li (1994) is consistent with the value required
to obtain the terminal velocity.

We assume for the purpose of solving the
equations of motion that $\log(U)$ stays the same throughout the flow.
Assuming the ionizing photons are isotropic, we can express the
density as $n=Q/(4\,\pi\,r^2\,U\,c)$, where $Q$ is the number flux of
ionizing photons.

Next, we find the filling factor $\epsilon(r)$ from the continuity
equation.  For convenience, we parameterize the outflow rate in terms
of the mass accretion rate, which we obtain from $L=\eta\, $\.M$\,
c^{2}$, assuming that the efficiency of conversion from gravitational
potential energy to radiation, $\eta$, is 10\%.  Thus, \.M$_{acc}$ is
the accretion rate for gas reaching the last stable orbit.  We define
a parameter $\Gamma \tbond $\.M$_{wind}/$\.M$_{acc}$, the ratio of the
outflow rate to the accretion rate.  We note that the outflow rate can
be greater than the accretion rate ($\Gamma > 1$).  This just means
that not all of the gas accreting from infinity actually approaches
the last stable orbit and accretes onto the black hole; some is blown
out as a wind at a relatively large radius.

At this point, we can also apply our final physical constraint. We
found the best estimate of the column density of the flow to be
$\log(N_H)=21.4$.  We can integrate $\epsilon(r)\, n(r)$ over $r$ to
obtain the column density.  But first we consider the line profile.
Because we impose the constraint that the ionization parameter is
constant along the flow, the line emissivity per unit radius will be
proportional to $\epsilon\,n\,r^2$.  Along a single streamline, this
is a steeply decreasing function of $r>r_f$, as a consequence of the
continuity equation.  This means that most of the mass is at low
velocities; however, the profile we see is peaked not at zero velocity
but rather at $\sim 2,500 \,\rm km\,s^{-1}$.  This means either that
our admittedly simple velocity law is not sufficient to describe the
wind, or the gas at zero velocity must not be emitting significantly;
below, we discuss why the low-velocity gas might not be emitting.
Regardless, it means that to obtain the column density, we do not
integrate the flow from $r_f$ but rather from $r>r_f$ where $v$
is larger than zero.  We choose, rather arbitrarily, the lower limit
on the integral to be the radius at which the observed velocity
$v\approx 2,000\,\rm km\,s^{-1}$.   For the upper limit of the
integration, we consider that at some radius, the clouds finally give
in and evaporate. We choose the final radius to be that where the
observed velocity is $8,000\,\rm km\,s^{-1}$, which may be considered
to be the largest velocity at which a significant amount of emission
is seen.  Then, for a given $r_f$ and $l_{edd}=L/L_{edd}$, we can adjust
$\Gamma$ until the calculated column density matches the measured
value. 

There are a couple of ways that gas could conceivably be accelerated
from zero velocity in such a way that we will not see it initially in
the wind.  For example, the low-velocity gas could be accelerated by
the continuum, but not excited by it.  That would be true if the
spectrum incident upon the lowest velocity gas, presumably closest to
the gas source, is transmitted through gas that is optically thick at
the Lyman edge.  Such a continuum is a primary assumption in the model
developed by Proga, Stone \& Kallman (2000).  Then, as the gas reaches
higher and higher velocities, it is exposed to increasingly
less-absorbed continuum, eventually starting to emit.  Another
possibility is that the gas is accelerated by another mechanism than
line driving initially. One possibility is that the gas is initially
dusty, and this dusty gas is accelerated by radiation pressure on the
gas due to the radiation field from the accretion disk. Finally, it
could be that the low-velocity gas is the origin of the intermediate
ionization lines, and is dense enough to be characterized by a lower
ionization parameter.  Further discussion of these scenarios is left
to a future paper.

The final step is to check the consistency of the model.  We do this
by computing the effective optical depth $t$. Then, using this
value plus the ionization parameter $U$ obtained from the
observations, we obtain the force multiplier $M_L(U,t)$ from the
tables given in Arav \& Li 1994.  This computed force multiplier must
be consistent with the value required to accelerate the flow to the
observed terminal velocity of $10,000\rm\, km\,s^{-1}$.  We
immediately find that the values of the effective optical depth
parameter $t$ are on the order of $10^{-7}$ to $10^{-9}$.  For
$\log(U)=-0.8$, the tables in Arav \& Li 1994 give $\log(M_L(U,t))$ of
around $3.1$.  Finally, we consider the fact that we may not be
viewing the wind along a stream line, and thus the velocity that we
see is less than the velocity along the stream line by a factor of
$\cos(\Theta)$, where $\Theta$ is the angle between the normal and the
wind streamline.  For each value of $\cos(\Theta)$, we determine the
wind foot-point for the self-consistent solution, defined by the
observed ionization parameter, column density, and inferred observed
terminal velocity of $10,000\,\rm km\,s^{-1}$.  An example of the
result is shown for two values of $\cos(\Theta)$ in Fig.\ 11.

If the object is radiating at the Eddington luminosity, the inferred
black hole mass is small, so the force multiplier required to
accelerate the gas to the terminal velocity is correspondingly small
at small values of the footpoint.  Thus, the consistency requirement
described above is not met for radii smaller than $100,000 \rm \,R_S$.
This is a very large radius.  In contrast, for the larger black hole
mass, the force multiplier required to accelerate the gas to a
terminal velocity of $10,000\,\rm km\,s^{-1}$ reaches
$log(M_L(U,t))=3.1$ comfortably between 10,000 $R_S$, and 100,000
$R_S$ for a range of streamline angles with the normal, $\cos(\Theta)$.

While we can't rule out the small black hole solely from the
kinematics, we do note that this solution is not favored because it
implies that the emission region for the wind should be at a larger
radius than the emission region for the intermediate-ionization lines,
violating our assumption that the wind filters the continuum before it
illuminates the intermediate-ionization line-emitting region.  Also,
these radii are larger than the predicted radius for the breakup of
the accretion disk, where Collin \& Hur\'e (2001) propose that some of
the line emission is occurring.  Hereafter, therefore, we discuss only
the large black hole mass solution.

In Fig.\ 12, we plot the wind foot-point for the self-consistent
solution, defined by the observed ionization parameter, column
density, and inferred projected terminal velocity of $10,000\,\rm
km\,s^{-1}$ as a function of $\cos(\Theta)$.  We also plot the radius
at which the projected velocity 
reaches 5,000$\,\rm km\,s^{-1}$, $r_{5000}$, as well as other physical
parameters of the outflow, including density, filling fraction, and
fraction of the accretion rate in the outflow, multiplied by the
covering fraction 0.15 obtained by the photoionization modeling, for
the situation where $L/L_{Edd}=0.12$ and the X-ray weak continuum.  An
approximate measure of the radial extent of the outflow is the
difference between $r_f$ and $r_{5000}$ divided by $r_{5000}$.  This
parameter is found to be uniformly 0.25.  This implies that the wind
extends in a fine mist over a fair fraction of the inner tens of
thousands of Schwarzschild radii of the AGN.

\placefigure{fig12}

\section{Discussion}

\subsection{Summary of Inferences from Modeling}

The photoionization modeling and toy wind model presented in this
paper lead us to draw
several conclusions and inferences about the optical and UV emission
regions in IRAS 13224$-$3809 and 1H~0707$-$495.    In these two objects,
a continuum that is overall relatively soft and deficient in X-rays
illuminates gas with enhanced metallicity.  A low column-density wind
is accelerated that manifests itself through blueshifted
high-ionization emission lines including \ion{C}{4}, \ion{O}{4}],
Ly$\alpha$ and \ion{N}{5} (the wind component).  Representative physical
conditions in the wind derived from the photoionization modeling are
$\log(U)=-0.8$, $N_H=10^{21.4}\rm\, cm^{-2}$, $n=10^8 \rm\, cm^{-3}$,
and covering fraction 0.15.  Using a simple radiative line-driving
model, and assuming that we see the gas that is accelerated (i.e., no
invisible massive substrate is present), we find that the
photoionization results are consistent with the dynamical model if the
black hole mass is $1.3 \times 10^{8}\rm\, M_\odot$, and the radius of
foot point of the wind is $\sim 10,000\rm R_S$. We infer that the
object is radiating at about 12\% of the Eddington limit.

The continuum also illuminates higher density, lower velocity, higher
optical depth gas, producing narrow emission lines centered at zero
velocity (the disk component).  If the continuum illuminates this gas
directly, the ionization parameter cannot be higher than
$\log(U)\approx -3$.  However, if the continuum passes through the
wind before illuminating this gas, the resulting continuum, already
weak in hard X-rays, is now also deficient in EUV and soft X-rays
above the helium ionization edge at 54 eV.  The result is that the
high-ionization lines in the disk component are very weak, if present
at all, and thus the high-ionization lines in the spectrum are
dominated by the blueshifted wind components.  The disk component is
instead dominated by intermediate-ionization lines, including
Ly$\alpha$, \ion{Si}{4}, \ion{Al}{3}, \ion{Si}{3},\ion{C}{3}], and
\ion{Fe}{3}.  Prominent low-ionization lines, including \ion{Mg}{2},
\ion{Fe}{2}, and \ion{Si}{2} are also present. The line ratios in the
disk component indicate a slightly high density ($\log(n)=10.25$) but
we note that a shift toward relatively lower-ionization lines is
expected due to the relatively hard spectrum that the filtered
continuum presents (Casebeer \& Leighly 2004).  If the continuum is
filtered through the wind, the inferred ionization parameter is
$\log(U)=-2.1$, and the corresponding radius is $14,000 \rm
R_S$.  A relatively small covering fraction of $0.05$ is inferred.

\subsubsection{A Few Comments on the Robustness of Results}

We note that we do not claim that we can firmly rule out alternative
models on the basis of the results presented here.  The scenario
presented above was chosen because it appears to be most consistent
with the data.

We also caution that the models are very simple.  Thus, one may be
justified in asking, how well do they plausibly conform to the real
situation?  The ionization model for the wind may be quite generally applicable,
because we searched a very large region of parameter space: we varied
the density, column density, ionization parameter, and covering
fraction, as well as the metallicity and continuum shape to some
extent. Furthermore, we applied a figure of merit to quantitatively
gauge the applicability of our results.  Small improvements will be
attained by better constraints on the metallicity that will be
possible using the \ion{O}{6} emission (e.g., Hamann et al.\ 2002) in
the {\it FUSE} data (Leighly et al.\ in prep.).  The {\it FUSE} data
will also be very valuable in trying to determine whether there is any
ionization variation with velocity.

How good is the toy wind model?  The intent of the model was to simply
make an estimate of the distance to the wind.  If radiative-line
driving is the acceleration mechanism, then the magnitude of the
acceleration depends directly on the effective optical depth, which in
turn depends directly on the volume filling factor, and the density,
and inversely on the velocity gradient (Arav \& Li 1994).  As is, we
obtain the velocity from the conservation of momentum equation.  Then,
the gradient of the velocity scales with footpoint radius: a wind
starting close to the AGN will reach the terminal velocity on a small
size scale, implying a relatively compact emitting region and large
velocity gradient, while a wind starting far from the AGN will reach
terminal velocity on a larger size scale, implying a small velocity
gradient.  The other important constraint comes from the requirement
that integrating over the wind gives the same column density as that
inferred from the photoionization modeling.  Then, combining the size
scale with the column density gives the filling factor.  Since the
integrated column density must be the same no matter where the wind
starts, the filling factor and velocity gradient counterbalance each
other to make the effective optical depth constant.  Thus, the toy
model almost boils down to a simple Eddington limit problem; we have
determined the radius at which radiation pressure, including
radiative-line driving appropriate for the ionization state, velocity
gradient and filling factor, opposing gravity can accelerate the wind
to yield the inferred terminal velocity of $10,000/\cos(\Theta)\rm
km\,s^{-1}$.

How could this estimate be refined?  A more general velocity law that
is frequently used is the so-called beta velocity law $v(r)=v_{\infty}
(1-r_f/r)^\beta$, where $\beta$ is an adjustable parameter that is
equal to 0.5 in the case considered above.  The velocity gradient for
the beta velocity law still scales with the footpoint radius, so if we
make the same assumptions as above, that the integrated column is
required to be that inferred from the photoionization modeling, and
radiative line-driving provides the acceleration against gravity, the
resulting inferred radius for the emission region would probably not
be much different unless extreme values of $\beta$ were used.

The situation would change if we imagine that there is a significant
amount of gas that we do not see that is also being accelerated.  For
example, as the flow accelerates, some of the gas may become too
ionized to emit in the UV; such gas is currently not accounted for.  A
more massive flow requires a smaller footpoint, because a larger
radiation force is needed to accelerate it to the terminal velocity.
Another potential complication is that the streamlines could change
direction, so that the component of the velocity parallel to the line
of sight changes.  Many other complications could be imagined, and at
some point, dynamical models coupled with photoionization models are
needed. 

In summary, the photoionization modeling results appear to be quite
general, because we searched a large region of parameter space.  If
we believe that acceleration mechanism is radiative-line driving, and
that there is no heavy invisible substrate, we may be confident that
the radius estimation for the wind is fairly accurate.  It is not
possible to put specific confidence limits on the estimate without
much more work, which is beyond the scope of this paper.

\subsection{Comparison with Previous Results}

\subsubsection{Wilkes et al.\ 1999 \& Kuraszkiewicz et al.\ 2000}

Wilkes et al.\ (1999) investigate emission line strengths and widths
of 41 quasars and compare them with their continuum properties.  The
quasars are a heterogeneous group chosen because they had {\it
Einstein} spectra; therefore, there is an inherent bias against
objects with large values of $\alpha_{ox}$, as noted in their paper.
Also, their sample is comprised almost half of radio-loud objects.
Among other things, they find an anticorrelation between
EW(\ion{C}{4}) and $\alpha_{ox}$, as we do. In their sample, the
objects with the lowest infrared (0.1--0.2 $\mu$) luminosity are five
NLS1s and 2 broad absorption-line quasars.  These objects all have
relatively weak \ion{C}{4} lines; when removed from the sample, Wilkes
et al.\ recover a Baldwin effect for \ion{C}{4}.

Kuraszkiewicz et al.\ (2000) follow Wilkes et al.\ (1999) with a more
detailed analysis of the NLS1s in the Wilkes et al.\ sample,
supplementing those with a few additional objects.  They analyze some
of the same {\it HST} spectra that we did in Paper I, although they do
not look at IRAS~13224$-$3809 and 1H~0707$-$495.  They include {\it
IUE} spectra as well.  They do not attempt a velocity deconvolution of
the lines, but rather draw inferences from the total flux of the line;
certainly, the low resolution of the {\it IUE} data prevents detailed
profile studies.  Thus, they perform 1-zone {\it Cloudy} modeling of
the line flux, rather than modeling of two regions, as we do here,
although in the end they infer a stratified BLR.  They infer that a
density of $10^{11-12} \rm\, cm^{-3}$ is required to produce the weak
\ion{C}{4}, and that this density is about 10 times higher than that
inferred from reverberation mapping results ($10^{11}\rm cm^{-3}$;
Peterson et al.\ 1985).  They also infer a low ionization parameter of
$\log(U)=-3$.

As noted in \S 2.2, our Type I solution is quite similar to that
inferred by Kuraszkiewicz et al.\ 2000.  Thus, the origin of the
differences between our final solution and theirs apparently lies in
primarily the much larger region of parameter space that we explored.
Also, the velocity deconvolution of the line profile allows us to
examine the conditions of the intermediate- and high-ionization line
regions separately.  Kuraszkiewicz et al.\ (2000) assume solar
abundances; we explore a few cases of enhanced metal abundances. 
We also use \ion{N}{5} which is very important for constraining
abundances and the ionization parameter.  Kuraszkiewicz et al.\ (2000)
choose the spectral energy distribution from PG~1211$+$143, including
the infrared. This object appears to have strong X-ray flux and
therefore rather powerful heating.  As discussed in \S 2.2, the
abundances and X-ray flux are very important for determining
the survival of ions under high radiation fluxes.

\subsubsection{Richards et al.\ 2002}

Richards et al.\ (2002) recently reported analysis of quasar spectra
from the Sloan Digital Sky Survey (SDSS; York et al.\ 2000).  They
found that the high-ionization lines, in particular \ion{C}{4}, are
systematically blueshifted with respect to the low-ionization lines
\ion{Mg}{2} and [\ion{O}{3}].  They found an anticorrelation
between the shift of \ion{C}{4}, and its equivalent width, similar to
the result reported here.  Examination of the profiles of composite
spectra constructed according to \ion{C}{4} blueshift leads them to
interpret the shift/EW anticorrelation as a lack of red wing emission
rather than a real shift.  This is consistent with the interpretation
presented here and in Leighly (2001).  

We differ with Richards et al.\ (2002) on the interpretation of this
inference.  They propose that all quasars have the same
components (e.g., disk, wind, emission line regions), and the
anticorrelation is a consequence of differing orientations.
Specifically, they propose that blueshifted lines are seen in objects
that are edge-on, like a broad absorption line quasar (BALQSO), with
the difference being that the flow is not directly in the line of
sight.  The receding red side is blocked by some kind of screen.  They
find two pieces of supporting evidence for this hypothesis. First,
they find similarities in the spectra with the most strongly
blueshifted \ion{C}{4} (their Composite D) with low-ionization BALQSOs 
(loBals).  Thus, if all AGN have BAL winds, and loBALs are viewed at
an angle large with respect to the symmetry axis, then Composite D
objects may be viewed edge on.    However, there is another explanation.
Richards et al.\ (2002) note that both the loBALs and the Composite D
objects have weak \ion{He}{2}. This is classic evidence for weak soft
X-ray continuum emission.  Thus, it may be that both Composite D and
the loBALs are soft X-ray weak, allowing a strong, low-ionization wind
to form.  The second piece of supporting evidence is the fact that
there are significantly more radio-detected (FIRST survey) objects in
the composite with the least blueshift (Composite A) compared with the
objects with largest blueshifts (Composite D).  Since FIRST survey
will resolve out lobe emission, they interpret this as evidence that
the Composite D are edge-on.  However, another speculative explanation
could be that Composite A objects may have radio jets instead of
winds, and the base of the jet may provide additional illumination of
the disk component, producing stronger emission near the rest
wavelength.  

A final difference in interpretation comes for the 1400\AA\ feature
for which they note little difference between the composites.  The
analysis presented here shows that 1400\AA\/ is composed of
\ion{Si}{4} from the disk and \ion{O}{4}] from the wind.  It may be
that the shift of the spectra from disk-dominated to wind-dominated is
accompanied by a shift from \ion{Si}{4} to \ion{O}{4}] in a way that
maintains an approximately constant line flux and profile.

\subsubsection{Brotherton et al.; Wills et al.}

In a series of three papers, Wills and Brotherton explored the
emission line properties of a sample of high signal-to-noise restframe
UV spectra from 123 high-luminosity moderate-redshift AGN (Wills et
al.\ 1993; Brotherton et al.\ 1994a,b).  They found a number of
correlations that are similar to those found here.  For example, Wills
et al.\ (1993), in studying the \ion{C}{4} line and 1400\AA\/ feature,
found that as the width of the \ion{C}{4} increased, the equivalent
width, kurtosis, and peak-to-continuum intensity decreased.  We did
not compile the widths of \ion{C}{4}, but we can understand the
asymmetry parameter to be nearly equivalent because of our ability to
explain the equivalent width/asymmetry anticorrelation observed
with the sum of a strongly blueshifted wind component and a narrow,
symmetric component (Paper I).  They also see the structure of
the 1400\AA\/ feature change in a way that can be interpreted as
decrease in \ion{Si}{4} to \ion{O}{4}] ratio with the increasing width
of \ion{C}{4}.  This seems to be consistent with our idea that when
the 1400\AA\/ feature is dominated by disk, corresponding to narrow
\ion{C}{4}, it should be primarily composed of \ion{Si}{4}; when
dominated by wind, corresponding to broad \ion{C}{4}, the 1400\AA\/
feature should have a significant component of \ion{O}{4}].

Brotherton et al.\ (1994a) study the relationship between \ion{C}{4},
the 1900\AA\/ feature comprised primarily of \ion{C}{3}], and
\ion{Mg}{2}.  They find that as the FWHM of \ion{C}{4} increases, the
ratio of \ion{C}{3}] to \ion{C}{4} increases.  This is consistent with
our scenario, because when \ion{C}{4} is dominated by wind, and
therefore is measured to be broad, there is no contribution from the
disk in the \ion{C}{4} line, so \ion{C}{3}] appears to be relatively
strong. 

Based on these detailed analyses, (Brotherton et al.\ 1994b) propose
that the broad lines in active galaxies are composed of two
components: a component from an intermediate-line region that consists
of a relatively narrow (FWHM $\sim 2000\rm\, km\, s^{-1}$) line at the
systemic redshift, and a much broader (FWHM $\sim 7000\rm
\,km\,s^{-1}$), somewhat blueshifted component.  This is similar to
our proposed deconvolution.  However, the physical interpretation
differs.  They place the VBLR very close to the central engine,
whereas we place it rather far away from the central engine.  Another
difference seems to be that they don't consider the possible effect of
enhanced abundances or continuum shape. Also, they place the
intermediate-line region very far from the central engine, whereas we
show that it can be brought into a smaller radius by filtering the
continuum through the wind.  Interestingly, they derive similar
covering fractions for both regions as we do.

\subsubsection{Wills et al.\ 1999; Wills et al.\ 2000; Shang et al.\
2002}

In another series of three papers, Wills and coworkers report the results of
analysis of {\it HST} spectra from a sample of 22 quasars (Wills et
al.\ 1999; Wills et al.\ 2000; Shang et al.\ 2002). This sample is
drawn from a complete sample of 23 optically-selected PG quasars,
chosen to have low redshift and low Galactic hydrogen column
densities (also, Laor et al.\ 1997a).

Wills et al.\ 1999 present a first look at the spectra, and describe
an extension of the optical-X-ray Eigenvector 1 to the UV properties.
Narrow lines (including \ion{C}{3}]) are linked with a larger
\ion{Si}{3}]-to-\ion{C}{3}] ratio, stronger low-ionization lines,
weaker \ion{C}{4}, and stronger \ion{N}{5}.  These are just the
properties that describe the spectra of IRAS~13224$-$3809 and
1H~0707$-$495. This is not surprising, since these two objects are
Narrow-line Seyfert 1 galaxies and are sufficiently blue and point-like
to have been classified as PG quasars had they been located in the
surveyed portion of the sky, although, like I~Zw~1, they would not
have been included in the Wills et al.\ (1999) sample because their
Galactic column is too large.  Thus, the model that explains our
objects can also be applied to the interpretation of the UV extension of
Eigenvector 1: \ion{C}{4} is weak because it originates in a wind with
little contribution from a narrow, symmetric component.  \ion{N}{5} is
strong because abundances are high, but also possibly because there is
some Ly$\alpha$ pumping of \ion{N}{5} in the wind.  Low-ionization lines are strong
because the continuum is soft, and also perhaps because the wind
filters the continuum before it illuminates the low-ionization line
emitting region, leaving it no alternative but to cool by
low-ionization line emission.  Filtering and a soft continuum can also
influence the \ion{Si}{3}]-to-\ion{C}{3}] ratio.

More detailed analysis is presented in Wills, Shang \& Yuan (2000) and
especially Shang et al.\ 2002.  The UV spectra were combined with
optical spectra and a spectral principal components analysis was
performed over the unprecedented large wavelength range.  They find
that 79\% of the variance lies in the first three eigenvectors, but
this time the first eigenvector describes correlations between the
luminosity and the relatively narrow (FWHM$\approx
2000\rm\,km\,s^{-1}$) core of the line. This is interpreted as a
manifestation of the Baldwin effect (Baldwin 1977).  We discuss the
Baldwin effect in the context of our interpretation in \S 4.6.2.
The second eigenvector found by Shang et al.\ (2002) is associated
with the continuum slope, and may be influenced by reddening.  The
third eigenvector is associated with the traditional first eigenvector
discussed by Boroson \& Green (1992). It contains all of the classic
parameters associated with the Boroson \& Green (1992) Eigenvector 1
and the UV extension discussed by Wills et al.\ 1999.  An interesting
feature is that \ion{C}{4} is very broad in this eigenvector.

\subsection{Other Considerations for the Radius of the High-ionization Line Emitting
Region}

In \S 2.2, we showed that the wind emission lines are consistent with
three types of photoionization solutions: Type I, a high-ionization,
high-density solution; Type II, an intermediate-ionization,
intermediate-density solution, and Type III, a solution characterized
by a large range of densities.  We favor the Type III solution,
because it encompasses a large region of the photoionization parameter
space, and therefore seems to be the least fine-tuned.  But because a
large range of densities were allowed by the photoionization solution,
the distance from the central engine to the emission region could not
be constrained without the toy dynamical model presented in \S 2.2.
That analysis indicated a radius of $\ga 10,000 \,\rm R_S$ for a $1.3
\times 10^8\rm \, M_\odot$, depending somewhat on the angle the flow
streamline makes with the observer.  In this section, we comment on
the consistency of our large radius with reverberation mapping
results, take another look at the Type I solution, and investigate the
possibility of decreasing the radius of the high-ionization
emission-line region for the Type III solution by ``shielding'' the wind.

\subsubsection{The Type III Solution and Reverberation Mapping Results}  

The Type III solution combined with the toy dynamical model indicates
an emission region for the wind of $\ga 10,000 \rm \, R_S$ for a $1.3
\times 10^8\rm \, M_\odot$, depending somewhat on the angle the flow
streamline makes with the observer.  This corresponds to a distance of
$3.8 \times 10^{17} \rm cm$, or about 150 light days.  On the face of
it, an origin of the high-ionization line-emitting wind at such a
large radius seems to be in conflict with reverberation mapping
results, which find evidence for an ionization-stratified broad-line
region in which high-ionization lines are produced quite close to the
central engine.  For example, in NGC 5548, an object with UV
luminosity 4 times smaller than that of IRAS~13224$-$3809 and
1H~0707$-$495, the lag of the high-ionization lines behind the
continuum is on the order of 1 week or less (Korista et al.\ 1995).
In fact, there may be no conflict.  As discussed in Paper I, the
high-ionization lines in NLS1s can be considered to be comprised of
two components: the broad, blueshifted component that is produced by
the wind, and the narrow, symmetric component that is associated with
the disk.  Reverberation mapping results are probably dominated by the
narrow, symmetric core that should be produced at smaller radii.
Observations support this hypothesis, since in some cases the 
cores of the lines vary more than the wings (e.g., Wandel, Peterson \&
Malkan 1999).  This can be considered to be evidence that the
broad component is produced by optically thin gas, which is less
responsive to continuum changes (Shields, Ferland \& Peterson 1995).
However, it could also be produced if the wind were physically far
from the central engine, as is inferred here. 

\subsubsection{The Type I Solution Reconsidered}

In this section, we take another brief look at the Type I solution.
As discussed in \S 2.2, in this solution, the density is constrained
to very high values, $\log(n)\ga 11.5$.  The high density results in
partial thermalization of Ly$\alpha$ and \ion{C}{4} relative to
\ion{Si}{4}; thus, the observed low equivalent width of \ion{C}{4} is
explained.  This solution is quite similar to the result obtained by
Kuraszkiewicz et al.\ (2002), who analyzed UV spectra of NLS1s but
only considered the integrated line fluxes and did not analyze the
line profile.  The high density and high inferred ionization parameter
($\log(U)=-0.4$) imply a high photon flux ($\log(\Phi)=22.5$) which
then implies that the emission region is located quite close to the
central engine at $R=6.1 \times 10^{15}\rm\, cm$ for $H_0=50 \rm\,
km\,s^{-1}\,Mpc^{-1}$ and $q_0=0.5$ (our default choice of
cosmological parameters), or $2.7 \times 10^{15}\rm\, cm$ for
$H_0=70\rm\, km\,s^{-1}\,Mpc^{-1}$, $\Omega_M=0.3$, and
$\Lambda_0=0.7$. A radius of $6.1 \times 10^{15}\rm\, cm$ corresponds
to $160\,R_S$ for our larger black hole mass of $1.3 \times 10^8\,
M_\odot$.  This value is somewhat smaller, but the same order as the
inferred location of the disk wind proposed by Murray et al.\ (1995)
of $600\rm \, R_S$. 

It would be easy to differentiate between Type I and Type III
solutions from the variability of the high-ionization lines.  First,
the inferred radius for the Type I solution implies a light-travel
time from the central engine to the emission region of 1--2.4 days;
thus the emission lines could respond rapidly to changes in the
continuum. For the Type III solutions, no variability would be
expected on short time scales because the light travel time is large.
Furthermore, the Type I and Type III solutions are characterized by
different column densities and different continuum opacities.  Both
are optically thin to the continuum; however, the Type III solution is
thinner, so that while the highest ionization lines, \ion{O}{6} and
\ion{N}{5} should vary in response to the continuum, since they are
produced in front of the helium ionization front, \ion{C}{4} is
produced deeper in the photoionized slab, so that it should saturate
in response to flux changes (e.g, Shields, Ferland, \& Peterson 1995).
For the Type I solution, \ion{O}{6}, \ion{N}{5}, and \ion{C}{4} are
all produced near the illuminated face of the cloud, implying they
would all vary in response to the continuum. Finally, the dynamical
time scale for the Type I solution, assuming a $1.3 \times 10^8\,\rm
M_\odot$ black hole, is less than a year, so one might expect to also
see changes in the line profile.

There may be an additional problem with the Type I solution.  When the
densities are so high that the lines are thermalized, the cooling will
be dominated by continuum emission, such as the Balmer continuum
(Rees, Netzer \& Ferland 1989).  Neither IRAS~13224$-$3809 nor
1H~0707$-$495 show an especially strong Balmer jump (note that the
Balmer jump can be distinguished from the strong \ion{Fe}{2} by the
wavelength of onset; e.g.\ Dietrich et al.\ 2002). 

\subsubsection{Can Murray et al.\  Shielding Reduce the Radius of the
  High-ionization Line Emitting Region?} 

In 1995, Murray et al.\ suggested that a wind, driven by resonance
scattering, could have a relatively small footpoint ($\sim 600\, R_S$)
if there is also present a region of highly ionized gas ($U=10$) that
``shields''\footnote{As discussed in \S 3.2, we differentiate between a ``shielded''
continuum, which is assumed to have been transmitted through highly
ionized gas (e.g., Murray et al.\ 1995), and a ``filtered'' continuum,
which is assumed to have been transmitted through the wind while
ionizing and exciting it before illuminating the disk and producing
the observed intermediate- and low-ionization lines.} the wind from
the strongly ionizing soft X-ray emission 
from the central engine.  They then suggest that this wind could emit 
broad emission lines under the conditions that $U=1$--10.    Recalling
that our best-estimated ionization parameter for the Type III solution
is $\log(U)=-0.8$, this implies that this shielding would allow the
wind emission to be produced where fluxes are $100-1000$ times higher,
corresponding to radii more than 10 times smaller, assuming the same
density.   However, based on our experience with ``filtering'', as
applied to the disk emission lines, we suspect that this scenario is
untenable because there would not be sufficient high-energy photons
in the shielded continuum to create and excite the high-ionization
ions that we see in the wind.  In this section, we numerically test
the effect of shielding on the emission from the wind.

We devised the following numerical experiment.  We transmitted our
hard low-flux continuum through shielding gas having a metal abundance
and a nitrogen abundance a factor of 5 and 10 over solar,
respectively, and then use that transmitted continuum to illuminate
the wind.  The parameters describing the shielding gas are as follows:
we uniformly use $U=10$, and consider a range of column density from
$10^{21}$ to $10^{22.8}\rm \, cm^{-2}$.  The parameters describing the
wind are the following: it has uniformly $N_H^{max}=22$, and a range
of ionization parameters from $\log(U)=-1$ to $\log(U)=0.5$.  Note
that, as before, since we are interested in obtaining the radius of
the emission region, the $U$s that we quote are appropriate for the
unabsorbed continuum.

We determine a reference point for the simulations as follows.  The
transmitted continuum will be almost identical to the incident
continuum for the extreme minimum of shielding column density of
$10^{21}\rm\, cm^{-2}$, because for $U=10$, that gas will be almost
completely ionized.  Then, since the adopted value of $N_H^{max}=22$
and extreme minimum value of $\log(U)=-1$ for the wind are close to
the best wind parameters derived in \S 2.2, the results of this
combination of parameters will be close to the observed values.
Therefore, we adopt the results from this extreme combination of
parameters as reference values with which to compare the results for
larger values of shielding $N_H$, and larger values of wind ionization
parameter of $\log(U)$.  The reference points lie in the lower left
corner of each panel in Fig.\ 13.

We plot contours of the normalized equivalent width, defined as
the predicted equivalent width divided by the equivalent width
obtained for the reference set of parameters described above; thus,
contours marked by ``1.0'' terminate necessarily in the lower left
corner of each plot.  These contours show that as the shielding column
density increases, from left to right across each  plot, the highest
ionization lines decrease, as the photons required to create the
emitting ions are removed from the continuum.  At the same time, the
gas is still subject to an intense photoionizing continuum, and it
responds by increasing emission of lower ionization lines.  As the
radius decreases, the line emission decreases as ions become
over-ionized. 

\placefigure{fig13}

We interpret these results in terms of the Murray et al.\ (1995) and
Murray \& Chiang (1998) scenarios as follows: shielding by highly
ionized gas can certainly increase the emission of intermediate- and
low-ionization lines at higher fluxes than would be possible without
shielding.  However, shielding does not affect high-ionization lines
in the same way, because the photons absorbed out by the highly
ionized gas are those required to create ions such as N$^{+4}$
(I.P.=77 eV) and O$^{+5}$ (I.P.=113 eV).  Therefore, we conclude that
shielding cannot bring the wind closer to the nucleus.  We note that
this conclusion is not dependent on the column density in the wind.
We examined higher column densities and found that while the flux in
other lines increased, the \ion{N}{5} line contours stayed the same,
which means, not surprisingly, that with
$N_H^{max}=10^{22}\rm\,cm^{-2}$ we integrate through the entire 
N$^{+4}$ zone.

\subsection{Constraints on the Intermediate-Ionization lines}

In Paper I, we discovered several correlations among the equivalent
widths and ratios of the intermediate-ionization lines \ion{C}{3}],
\ion{Si}{3}] and \ion{Al}{3} in our sample of NLS1s.  Specifically, we
find that both the \ion{C}{3}]/\ion{C}{4} and the
\ion{Si}{3}]/\ion{C}{3}] ratios are correlated with \ion{Si}{3}] and
\ion{Al}{3} equivalent widths, but anticorrelated with the \ion{C}{3}]
equivalent width.  As discussed in Paper 1, these correlations could
potentially have an origin in a variation in density, ionization
parameter, or continuum shape.

To investigate quantitatively the complex interdependencies of
these three parameters, we run some {\it Cloudy} models using the
hard low-flux continuum.  We examine the equivalent widths of
\ion{C}{4}, \ion{Al}{3}, \ion{Si}{3}], and \ion{C}{3}] as a function
of density for a constant covering fraction of 0.05, and two
ionization parameters ($\log(U)=-2$ and $-3$). We assume that the gas
is ionization-bounded.  We compute the equivalent widths under two
assumptions: the continuum illuminates the emitting gas directly; and
the continuum illuminates the gas after passing through the wind, as
discussed in \S 2.3.  The results are displayed in Fig.\ 14.

\placefigure{fig14}

Fig.\ 14 shows the decrease in equivalent width as a function of
density expected for the semiforbidden lines \ion{C}{3}] and
\ion{Si}{3}].  The permitted line \ion{C}{4} shows little dependence
on density, while \ion{Al}{3} shows a moderate increase, possibly due
to its increasing role in cooling in the \ion{C}{3}] and \ion{Si}{3}]
forming region.  This seems to suggest that if an increase in density
drives the correlations, the \ion{C}{3}] to \ion{C}{4} ratio should
decrease, contrary to the observations.

Filtering the continuum through the wind results in a dramatic
decrease in \ion{C}{4}.  In fact, there is a larger decrease between
the filtered and nonfiltered continua for $\log(U)=-2$ than there is
between either continuum with $\log(U)=-2$ and $\log(U)=-3$.  For
$\log(U)=-2$, there is little effect on the intermediate-ionization
lines \ion{Al}{3}, \ion{Si}{3}] and \ion{C}{3}], although we see a
slight increase for \ion{Si}{3}] for the filtered continuum.  For
$\log(U)=-3$ and a filtered continuum, all the lines become weaker,
with \ion{C}{4} decreasing the most.

These simple simulations show that the origin of the correlations
among the intermediate ionization line properties could be a filtered
continuum, or it could originate from a trend 
in ionization parameter, but an origin in density variations does not
seem to be supported.

\subsection{Ideas and Speculations about the Emission
  Regions in NLS1s}

In this paper and in Paper I, we learned many things about the UV
emission lines from two particular NLS1s, 1H~0707$-$495 and
IRAS~13224$-$3809, and about the properties of those two objects
compared with a heterogeneous sample of NLS1s.  In this section, we
attempt to use these results to construct a scenario for the emission
regions in NLS1s. 

Perhaps the simplest scenario for the emission-line geometry
consists of an accretion disk with a wind.  The blueshifted part of
the lines is produced in the wind, while the portion centered at the
rest wavelength is produced in the base of the wind, or in the
accretion disk itself.  In some objects, there is no significant wind,
in which case all the line emission comes from the base of the wind.
In others, the wind emission is relatively strong; in these objects,
high ionization lines have a blue wing arising from the wind.

Why do some objects have winds and others do not?  If the winds are
accelerated by radiation-line driving, then objects with steep
$\alpha_{ox}$  may have winds, while objects with flat $\alpha_{ox}$
may not. This is because a strong UV continuum is necessary to drive
the wind; however, too much soft X-ray emission will easily 
overionized the gas, destroying the resonance-scattering ions.   A high
metallicity may also encourage a wind, since that could potentially
contribute more resonance-scattering ions.  

This is the simplest scenario; however, it does not explain a number
of aspects of the data.  It does not explain why \ion{C}{4} is so weak
in the wind-dominated NLS1s like IRAS~13224$-$3809 and 1H~0707$-$495;
the scenario above only allows blue wings; the \ion{C}{4} line core,
and other high-ionization line cores,  produced in the base of the
wind should still be very strong, and should dominate the line
emission.  It also does not explain the extreme line ratios in the
wind-dominated NLS1s, including relatively large
\ion{C}{3}]/\ion{C}{4},   \ion{Si}{3}]/\ion{C}{3}], and
\ion{Al}{3}/\ion{C}{3}] ratios.   Additional tweaks of the scenario
are necessary to explain these properties.

How to explain the weak \ion{C}{4} core in the wind-dominated NLS1s?
Increasing the density will not help.  Although the high
\ion{Si}{3}]/\ion{C}{3}] ratio suggests a higher density in these
objects, the inferred density, well constrained by the
\ion{Si}{3}]/\ion{C}{3}] to be $\log(n)\approx 10.25$, is not nearly
high enough to suppress \ion{C}{4}, a permitted line.  This point was
discussed in detail in \S 4.4.

Another possibility is to simply move the emission region very far
from the nucleus such that the dominant ionization state for carbon in
the base of the wind is C+2.  This possibility is explored in the
context of photoionization modeling in \S 2.3.  There are two
potential objections to this.  First, the inferred distance to the
emission region is very large, so that for reasonable black hole
masses, the Keplerian velocities are quite low.  A more important
problem would be, why is the emission region exceptionally far from
the nucleus in the wind-dominated NLS1s alone?  One possible answer
could be to associate the emission region with the breakup radius of
the accretion disk (Collin \& Hur\'e 2001), as that should vary with
accretion rate (Hur\'e 2000).  However, this does not simply solve the
problem, since the breakup radius depends inversely on accretion rate;
the larger the accretion rate with respect to Eddington, the smaller
the radius, just the opposite dependence required if we associate the
wind-dominated NLS1s with a higher accretion rate. 

In \S 2.3 we present a scenario that explains these observational
results naturally.  If the continuum is transmitted through the wind
before it illuminates the intermediate- and low-ionization line
emitting gas, it will lack sufficient photons in the helium continuum
to excite high-ionization lines.  Thus, the presence of a wind
naturally results in a weak core for the high-ionization lines.  The
softening of the continuum also tends to create and excite ions with
lower ionization potentials; this can partially explain the dominance
of \ion{Si}{3}] and \ion{Al}{3} in the intermediate-ionization
line-emitting region.

\subsubsection{A Speculative Scenario}

The analysis presented in this paper, and in Paper I, leads us to
infer that the spectral energy distribution is a primary factor in
determining both the ionization and the dynamics of the line-emitting
gas in AGN.  A continuum strong in the UV and weak in X-rays can drive
a wind by resonance scattering without overionizing the ions required
for the line driving; this results in blueshifted high-ionization
lines.  This wind may then filter the continuum before it illuminates
lower-velocity gas emitting intermediate- and low-ionization lines.
The filtered continuum lacks high energy photons, so the low-velocity
gas cools by emitting particularly lower-ionization lines, including
such low-ionization lines as \ion{Fe}{2} and \ion{Si}{2}.  Conversely,
if the X-rays are strong compared with the UV, a wind is not formed
because the gas is overionized before it can be accelerated.  This
unfiltered continuum illuminates the lower-velocity gas directly.
This continuum is strong in the extreme UV 
and in X-rays, so higher-ionization lines, including \ion{C}{4}, are
produced in the low-velocity gas.

The scenario outlined above begs the question: what determines the
continuum to begin with?  The continuum may be directly controlled by
intrinsic parameters such as the black hole mass and accretion rate,
but how?  We present a speculative scenario to explain this as
follows.  Under conditions of moderate accretion rate, the accretion
disk may be ``thin'', at least at large radii (e.g., Frank, King \&
Raine 1992).  The thin disk is able to radiate the the power that it
generates, is optically thick and geometrically thin, and its spectrum
is predicted to be $F(\nu) \propto \nu^{1/3}$.  Increasing the
accretion rate causes the characteristic temperature to increase;
increasing the black hole mass causes the characteristic temperature
to decrease (e.g., Ross, Fabian \& Mineshige 1992).  Thus, the thin
disk can produce a blue continuum in the optical and UV that is harder
for smaller black hole masses and higher accretion rates.

At smaller radii, the thin disk is predicted to generate more energy
than can be radiated through its surface.  Radiation pressure should
become important, causing the disk to puff up and become geometrically
thick.  The simplest radiation-pressure dominated disks are known to be
unstable; however, detailed models show that this solution may be
still viable  (e.g., Agol et al.\ 2001; Blaes \& Socrates 2001).

X-rays are thought to be produced close to the black hole by an
optically thin, hot corona that intercepts optical/UV photons and
upscatters them.  The amount of X-ray emission depends on the amount
of corona; specifically, the covering fraction and optical depth.  The
amount of corona may be depend on how much accretion energy is
diverted into producing the corona, by, for example, magnetic
reconnection of buoyant loops generated in the disk that escape
through the surface.  Some recent models predict that the amount of
energy produced in the corona should also depend upon accretion rate
(e.g., Liu et al.\ 2002), so that high accretion rate objects have a
smaller fraction of energy emitted in the corona compared with the
disk. This scenario was used by Bechtold et al.\ (2003) to  explain
the steepening of $\alpha_{ox}$ with luminosity. 

So, combining the thin disk, the radiation-supported thick disk,  and
the corona, we suggest a speculative geometrical scenario that is
illustrated in Fig.\ 15 (not to scale).  The key feature that
potentially explains the behavior of the UV emission lines in NLS1s is
the spectrum emitted by the outer part of the geometrically thick
disk, because that is the spectrum that initially illuminates the
line-emitting gas.  If that continuum is dominated by photons
below 13.6 eV, it can accelerate a wind without overionizing it.  That
accelerated wind, which may originate in the region where the disk
becomes gravitationally unstable, reaches a modest velocity, yet does
not emit very much line emission.  As it rises above the lip of the
radiation-pressure supported torus, it is exposed to the full UV
continuum, is quickly accelerated and emits.  An analogy may be wind
blowing snow off a cornice.  Beyond the wind may lie the clumps from
the gravitationally unstable disk that would emit the lower-velocity, 
intermediate and low-ionization lines  when they are illuminated by
the continuum filtered through the wind.  

\placefigure{fig15}

In contrast, if the spectrum of the outer part of the geometrically
thick disk emits very much radiation at energies above 13.6 eV, the
potential reservoir for the wind gas is exposed to the full ionizing
continuum and is overionized before it can be accelerated.  
Furthermore, the thin disk will be illuminated by a strong
photoionizing continuum, and it will cool by emitting strong permitted
lines including \ion{C}{4}. 

This speculative scenario explains the emission lines, but it can also
explain the continuum.  Quasars that have strong blueshifted lines
tend to have blue optical and UV continua.  This may be a consequence
of the optical/UV emission from the thin disk dominating to a small
radius.  An example of such an object is PHL~1811, a very luminous
quasar (Leighly et al.\ in prep.)  Conversely, objects that do not
have blueshifted high-ionization lines may have less contribution from
the optically thick, geometrically thin disk; their continuum is
stronger in the extreme UV and may peak in the soft X-rays.  An
example of such an object is RE~1034$+$39 (Casebeer \& Leighly 2004).

\subsection{Other Considerations}

\subsubsection{Influence of X-ray Properties}

We have noted already that IRAS~13224$-$3809 and 1H~0707$-$495 have
extreme X-ray properties compared with other NLS1s, as discussed in
Leighly 1999b.  These properties include the highest-amplitude X-ray
variability and the most prominent X-ray soft excesses in their {\it
ASCA} spectra.  As discussed above, the photoionization modeling shows
that a larger region of parameter space is available when the X-ray
flux is low.  We justified our assumption of low X-ray flux using the
fact that 1H~0707$-$495 has been frequently found in very low flux
states, a factor of 10 lower than observed when the {\it ASCA}
spectrum was made (Leighly et al.\ 2002; Leighly et al.\ in prep.).
Furthermore, in Paper I, we showed that there is an anticorrelation
between the asymmetry of the emission line and $\alpha_{ox}$, such
that X-ray weak objects have blueshifted profiles.   

It is also possible that the high-amplitude variability plays a role
in the production of the emission lines.   If the high-amplitude
variability extends from the X-ray into the UV, a variable radiation
line-driving force may be present, or a variable photoionizing
continuum.  A variable line-driving force may be interesting because
it could produce shocks in the wind that may result in density
enhancements and may help formation of filaments dense enough to emit
the lines that we see.  A variable photoionizing continuum can effect
the ionization balance of the emitting region; Nicastro et al.\ (1999)
found that the effect was such that the emission region was
overionized compared with ionization expected for the particular
ionization parameter.  

There is no data yet that supports rapidly variable UV emission in
these two objects.  Young et al.\ 1999 do not see any optical
variability on short time scales over three nights in
IRAS~13224$-$3809.  No significant variability in the near-UV was
detected during a 20,000 second {\it XMM-Newton} OM observation.
However, observed correlated optical and X-ray variability in NGC~5548
led Uttley et al.\ (2003) to propose that relatively luminous Seyferts
may be more variable in the UV than less luminous objects because of
relatively cooler accretion disk expected for a larger black hole and
corresponding shift of the UV emission region to smaller $R/R_S$ where
dynamical time scales may be short. Thus, variability in the
line-driving continuum or photoionization continuum may be possible.

\subsubsection{The Baldwin Effect}

It is well known that the equivalent widths of emission lines in
quasars are anticorrelated with their luminosity (the Baldwin effect;
The Baldwin 1977; see Osmer \& Shields 1999 for a review).  Baldwin effect
has been variously attributed to variations in the broad-line cloud
covering fraction, ionization parameter, and inclination.  Perhaps the
most promising explanation is in terms of the shape of the continuum,
because it can explain the fact that the higher-ionization lines
experience a steeper decrease with luminosity than the
lower-ionization lines.  Another interesting feature of the Baldwin
effect is that it is stronger in the line cores. This fact was also
observed in the spectral PCA analysis of the {\it HST} sample of PG
quasars (Shang et al.\ 2002).

It is possible that our scenario can contribute to the interpretation
of the Baldwin effect.  Luminosity may be correlated with the black
hole mass, predicting a softer spectrum for larger black holes for a
fixed accretion rate relative to Eddington (e.g., Ross, Fabian \&
Mineshige 1992).  Thus, more luminous objects, because of their softer
spectrum, may produce a wind that may filter the continuum, reducing
the availability of energetic ionizing photons to the low-velocity
gas, reducing the line core.  Interestingly, Shang et al.\ (2002) find
that broad \ion{He}{2}$\lambda 4686$ is present in their first
eigenvector that they interpret as having an origin in the Baldwin
effect, and is correlated with the strength of the narrow component of
the lines.  As discussed above, \ion{He}{2} is the classic indicator
of a continuum strong in soft X-rays.  It is also interesting to note
that this interpretation contrasts with that of Shields, Ferland \&
Peterson (1995); they speculate that the covering fraction of the
optically-thin component depends inversely on luminosity.

\subsubsection{BALQSOs}

In this paper, we show that a number of the emission line
properties of NLS1s are plausibly related to the presence or absence
of a radiatively-driven wind.  The other class of active galaxy in
which there is good evidence for a radiatively-driven wind is the
class of broad absorption-line QSOs (BALQSOs).  Of course, there is an 
important difference between BALQSOs and NLS1s: in BALQSOs, the wind
is in the line of sight to the nucleus, causing sometimes spectacular
absorption lines to be imprinted upon the spectrum, while in NLS1s,
the wind is not in the line of sight.  This difference could be simply
a matter of orientation.  One sees emission lines regardless of the
viewer orientation (disregarding anisotropic emission, for the
moment), so if the wind has some effect on the emission line
properties of NLS1s, then it may produce the same effects on the
emission lines in BALQSOs. 

Weymann et al.\ (1991) present a comprehensive comparison between the
emission-line properties of BALQSOs and non-BAL quasars.  As discussed
in the introduction to that paper, their work was motivated by various
reports of the following differences between BALQSOs and ordinary
quasars: 1) \ion{Fe}{2} is stronger; 2) \ion{Al}{3} is stronger; 3)
\ion{C}{4} is weaker; 4) \ion{N}{5} is stronger.  We note that these
properties describe the differences between NLS1s and ordinary
broad-line quasars.  Weymann et al.\ (1991) then present a very
careful and conservative analysis of samples of BALQSOs and
non-BALQSOs. While they conclude that, excluding the low-ionization
BALQSOs\footnote{Other authors have remarked upon the similarity
between NLS1s and the low-ionization BALQSOs, objects which have very
strong \ion{Fe}{3} and \ion{Fe}{2} (e.g.\ Leighly et al.\ 1997;
Boroson \& Meyers 1992, for the case of I~Zw~1).}, the emission lines
are very much the same in BALQSOs and non-BALQSOs.  However, a few
subtle differences remain.  There is an enhancement of \ion{N}{5}, and
there is an enhancement around \ion{Al}{3}, which they attribute to
\ion{Fe}{2} or \ion{Fe}{3}. They also find a correlation between
\ion{Fe}{2} and ``balnicity'' index, a measure of the BAL-ness of a
quasar.  In our scenario, these properties would be explained as
follows.  The \ion{N}{5} enhancement arises in the BAL flow, and it is
perhaps enhanced by scattering (as noted by Weymann et al.\
1991). However, we also speculate that the strong low-ionization lines
result from illumination by a continuum filtered through the BAL flow.
We point out that in our scenario, in general the lines such as
\ion{C}{4} are composed of both wind and disk components, and thus
differences between objects with and without winds may not be very
spectacular.

If the radiatively-driven winds in BALQSOs are related to the winds
that we discuss in this paper, we  might also expect that BALQSOs
should be deficient in X-ray emission.  Measuring the X-ray emission
from BALQSOs is difficult, because the X-rays are usually absorbed,
and it is difficult to deconvolve the spectrum and the absorption
robustly without making assumptions about the continuum shape.
Sometimes, good evidence for normal spectral energy distributions have
been found (e.g., Gallagher et al.\ 2002) .  However, there is at
least one instance in which evidence that the BALQSO is intrinsically
X-ray weak has been found (Sabra \& Hamann 2001).

\subsubsection{Early Universe Counterparts of NLS1s}

Recently, several investigators have sought the early universe
counterparts of NLS1s (e.g., Mathur 2000).  This is important, because
now a significant number of quasars with redshifts $z>4$ have been
discovered.  This epoch corresponds to $>90$\% of the age of the
universe; therefore, these quasars are young.  The accretion rate in
these objects is of special interest.  Luminous quasars must have
large black holes, and in order to grow large in such a short time,
they ought to be accreting at a rapid rate.

The properties of the objects discussed in this paper may give us a
hint about what kind of objects at high redshift correspond to
Narrow-line Seyfert 1 galaxies.  There are some objects that have been
found that have extremely strong and narrow emission lines
(e.g., Constantin et al.\ 2002).  The lines in these objects have huge
equivalent widths; clearly the emitting region is very well
illuminated.  If we consider the fact that the Baldwin effect slopes
for NLS1s typically lies significantly below that of other quasars
(Paper I; Wilkes et al.\ 1999), then that argues against these
high-equivalent width objects being early-universe counterparts of
NLS1s.

On the other hand, several high-redshift objects have been discovered
in the Sloan Digital Sky Survey that have very blue continua and no
apparent emission lines (Fan et al.\ 1999; Anderson et al.\ 2001).
One of the properties of NLS1s is their relatively low line equivalent
width and frequently blue continua.  Thus, the line-less quasars seem
much more similar to the high-luminosity NLS1s, and therefore they may
be the sought-after early-universe counterparts (Leighly, Halpern, \&
Jenkins 2004; Leighly et al.\ in prep.).

\section{Summary}

\begin{itemize}
\item This is the second in a series of two papers devoted to
understanding UV line emission in NLS1s.  In the first paper (Leighly
\& Moore 2004), we present the results of analysis of {\it HST} STIS
spectra of the extreme NLS1s IRAS~13224$-$3809 and 1H~0707$-$495.
Those spectra reveal broad, blueshifted high-ionization emission
lines, and intermediate- and low-ionization lines that are narrow and
centered at the rest wavelength. We interpreted this as evidence that
supports a scenario in which the high-ionization lines are emitted in
a wind, and the intermediate- and low-ionization lines are emitted in
the accretion disk or base of the wind.  We develop a template profile
for the wind from the \ion{C}{4} line, then model the strongest lines
in the spectra using the wind template, and a narrow, symmetric
component.  We found that the following lines are dominated by wind
emission: \ion{C}{4}, \ion{He}{2}, \ion{N}{5}.  These lines are
dominated by disk emission: \ion{Al}{3}, \ion{Si}{3}], \ion{C}{3}],
\ion{Mg}{2}.  Ly$\alpha$ and the 1400\AA\/ feature composed of
\ion{Si}{4} and \ion{O}{4}] have both disk and wind components.

\item The photoionization code {\it Cloudy} was used with the results
of Paper I to determine the conditions of the line-emitting gas.  For
the wind, we investigated the effects of a large range of density,
photoionizing flux, a parameter related to column density
($\log(N_H^{max})=\log(N_H)-\log(U)$), and covering fraction, and a
few  combinations of metallicity and continuum shape.  We
compared the simulation results with the data using a figure of merit,
defined to be the absolute value of the difference between the log of
the simulated and observed equivalent widths.  We found evidence for
three minima in parameter space.  The Type I solution, obtained for
every combination of metallicity and continuum, was characterized by a
very high density ($\log(n) \ge 11.5$, high ionization parameter
($\log(U) \ge -0.6$), and large column density ($\log(N_H^{max})\ge
23$).  This is similar to the result obtained by Kuraszkiewicz et al.\
(2000) for NLS1s without profile deconvolution.  When the metallicity
was enhanced and the continuum was X-ray weak, the Type III solution,
characterized by a large range in density ($-7 < \log(n) < 11$), high
ionization parameter ($-1.2 < \log(U) < -0.2$), and low column density
($\log(N_H^{max}) \approx 22.2$), was obtained.  A much larger region
of parameter space is encompassed by the Type III solution; therefore
it appeared to be less fine-tuned and more physically realistic than
the Type I solution.

\item Photoionization modeling for the disk lines was performed using
the results from the photoionization modeling of the wind lines; that
is, metallicity was assumed enhanced, and an X-ray weak continuum was
used.  The ratio of the intercombination lines \ion{Si}{3}] and
\ion{C}{3}] lines provided a good density constraint ($10 < \log(n) <
10.5$).  The low limit on a disk contribution of \ion{C}{4}, and
corresponding low ratio of \ion{C}{4} to \ion{C}{3}] requires a low
value of the photoionizing flux ($\Phi=17.5$), and a very large radius
for the emission region ($1.9 \times 10^{18} \rm \, cm$).  However, if
the continuum is ``filtered'', or transmitted through the wind before
it illuminates the intermediate- and low-ionization line emitting
region, the photoionizing flux can be higher ($\Phi=18.7$) and the
emission region radius smaller ($5.3 \times 10^{17} \rm\, cm$).  The
difference is that the filtered continuum lacks photons in the helium
continuum, so it is not able produce \ion{C}{4} as efficiently.  Such
a filtered continuum naturally produces the strong lower-ionization
lines seen in NLS1s.

\item We used the photoionization results to infer the location of the
emission regions.  We used two candidate black hole masses: a small
one, obtained assuming that the object was radiating at the Eddington
limit, and a larger one ($1.3 \times 10^{8}\rm\, M_\odot$; $L=0.12
L_{Edd}$), obtained directly from the photoionization modeling results
for the intermediate-ionization emission lines.  We assumed the Type
III wind solution, which by itself provides little constraint on
density and therefore no constraint on radius.  To obtain a radius
constraint for the wind, we used a simple dynamical model in which the
wind is accelerated by radiative-line driving.  We find that the
larger black hole mass is better accommodated by the dynamical model
results.  Representative values of the parameters inferred for the
wind are a density of $\sim 10^{8.5}\rm\,cm^{-3}$, a radius of $\sim
10^{4}\,\rm R_S$, a filling fraction of $\sim 10^{-5}$, and a wind
outflow rate about half that accreted into the black hole.

\item We discuss the results of both papers and compare with
previously published results.  The discovery, presented in Paper I,
that blueshifted line profiles are associated with steep
$\alpha_{ox}$, seems to provide evidence for radiative-line driving as
the wind acceleration mechanism.  The concept of filtering the
continuum through the wind before it illuminates the intermediate- and
low-ionization line emitting gas may help explain the very strong
low-ionization line emission seen in many NLS1s and some other windy
objects such as BALQSOs.  We also provide a speculative scenario that
may  explain the range of UV emission line properties in NLS1s. 
\end{itemize}



\acknowledgments
KML thanks many people for useful discussions, including Eddie Baron,
Mark Bottorff, Mike Brotherton, Martin Gaskell, Kirk Korista, Norm
Murray, Daniel Proga, Daniel Savin, Joe Shields and especially Jules
Halpern and the OU AGN group (Darrin Casebeer, Chiho Matsumoto, \&
Larry Maddox). KML thanks Gary Ferland for help using {\it
  Cloudy}. Support for proposal 
\# 7360 was provided by NASA through a grant from the Space Telescope
Science Institute, which is operated by the Association of
Universities for Research in Astronomy, Inc., under NASA contract NAS
5-26555. This research has made use of the NASA/IPAC Extragalactic
Database (NED) which is operated by the Jet Propulsion Laboratory,
California Institute of Technology, under contract with the National
Aeronautics and Space Administration. This research has made use of
data obtained from the High Energy Astrophysics Science Archive
Research Center (HEASARC), provided by NASA's Goddard Space Flight
Center. KML gratefully acknowledges additional support by NASA grant
NAG5-10171 (LTSA).



\appendix

\section{Further Comments on the Black Hole Mass and Accretion Rate}

In \S 3, we make two estimates of the black hole mass.  The first
assumed the luminosity is Eddington, simply because the Eddington
luminosity is a fiducial parameter of the system.   This estimate was not favored
because it predicted a too low velocity for the \ion{C}{3}]
emitting region (\S 3.1), and because the mass was so low
compared with the luminosity that the force multiplier required to
accelerate the wind to the observed terminal velocity was lower than
implied by the gas properties (\S 3.2).  The second estimate was
made based on the calibration of the photoionization and reverberation
techniques described by Wandel, Peterson \& Malkan (1999), although we
did not make precisely the same assumptions as they did (see below).
The second estimate could indeed explain the width of the \ion{C}{3}]
line, not surprisingly, since that is built into the assumptions used
in the calculation.  Also, we could get a reasonable match between the
force multiplier required to accelerate the gas to the terminal
velocity, and the force multiplier defined by the gas properties.
We then infer that the object is radiating at 12\% Eddington
luminosity, and the photoionizing continuum is 8\% of the Eddington
luminosity.   This is comparable to that estimated for other objects
analyzed by  by
Wandel, Peterson \& Malkan (1999).  

However, Wandel, Peterson, \& Malkan (1999) would not have obtained
the same results as we have because of several different assumptions.
First, they assumed uniformly $Un_e=10^{10}$, uniformly, although it
is not clearly explained why this value is used. We use the values
inferred from the photoionization analysis for the conditions in the
\ion{C}{3}] emitting region: $U=0.008$ and $n_e=10^{10.25}
\rm\,cm^{-3}$.  These yield $Un_e=10^{8.2}$, a number that is much
smaller than that assumed by Wandel, Peterson \& Malkan (1999).  The
lower value of $Un_e$ that we use increases the inferred mass.
Secondly, we use the velocity width of \ion{C}{3}], whereas they use
the velocity width of H$\beta$, which is typically narrower.  In
IRAS~13224$-$3809 we measured H$\beta$ FWHM of $800\rm\, km\,s^{-1}$,
although we note that this could be contaminated by H$\beta$ from the
starburst, and therefore the width underestimated.  The larger value
of the velocity that we use increases the inferred mass.  Recomputing
the photoionization mass using $Un_e=10^{10}$ and $v_{FWHM}=800\rm \, km\,s^{-1}$ gives
a photoionization mass of $M_{ph}=2.4 \times 10^6 \,\rm
M_\odot$. Using their relationship between photoionization and
reverberation black hole masses gives $M_{bh}=4.3 \times 10^6
\rm\,M_\odot$.  This is lower than the estimate based on the
assumption that the object is radiating at the Eddington luminosity,
implying that it is super-Eddington.

How can we reconcile these differences? A number of factors may
influence our mass estimate.  As mentioned in \S 3.1, the
intermediate-ionization line-emitting region may see a lower flux than
we do 
because of $\cos(\theta)$ dependence of the emission (Laor \& Netzer
1989). It may be that the velocities in the intermediate-ionization 
line-emitting region are higher than the Keplerian velocity due to the
black 
hole alone because of the gravitational potential presented by the
heavy accretion disk (Hur\'e 2002).  Both of these factors would
decrease the estimate of the black hole mass required by the
intermediate ionization disk lines.

On the other hand, it may be preliminary to apply the reverberation
mapping results in bulk to all NLS1s.  Only a couple of the objects
that have been heavily monitored and are considered by Wandel,
Peterson \& Malkan (1991) have optical emission line properties that
securely classify them as NLS1s (NGC 4051 \& Mrk 335); Mrk~110
is better classified as a low-luminosity Seyfert 1.5.  Given
the range of UV properties exhibited by NLS1s, as discussed in Paper
I, is this too few on which to conclude the behavior of entire class?
It is also possible that some of the conclusions obtained from the
reverberation mapping results are compromised by selection effects.
It is well known that the Seyfert-luminosity objects that have been
intensively monitored were chosen because their Balmer lines were
already known to be variable; the quasars, on the other hand, were
chosen without such bias.  Kaspi et al.\ (2000) report that in the 
reverberation-mapped sample, there is an anticorrelation between
luminosity and Balmer emission line velocity width; this is opposite
the expectation of theoretical arguments (e.g., Laor 1998) and
observed in other quasar samples. 

Finally, we note that this difficulty in reconciling the results of
photoionization modeling with the small black hole masses inferred
from reverberation mapping (and in agreement apparently with galactic
bulge masses as well) may be similar to the apparent difficulty in
producing the optical continuum described by Collin \& Hur\'e (2001)
and Collin et al.\ (2002).





\clearpage







\clearpage

\clearpage

\begin{figure}
\epsscale{1.0}
\plotone{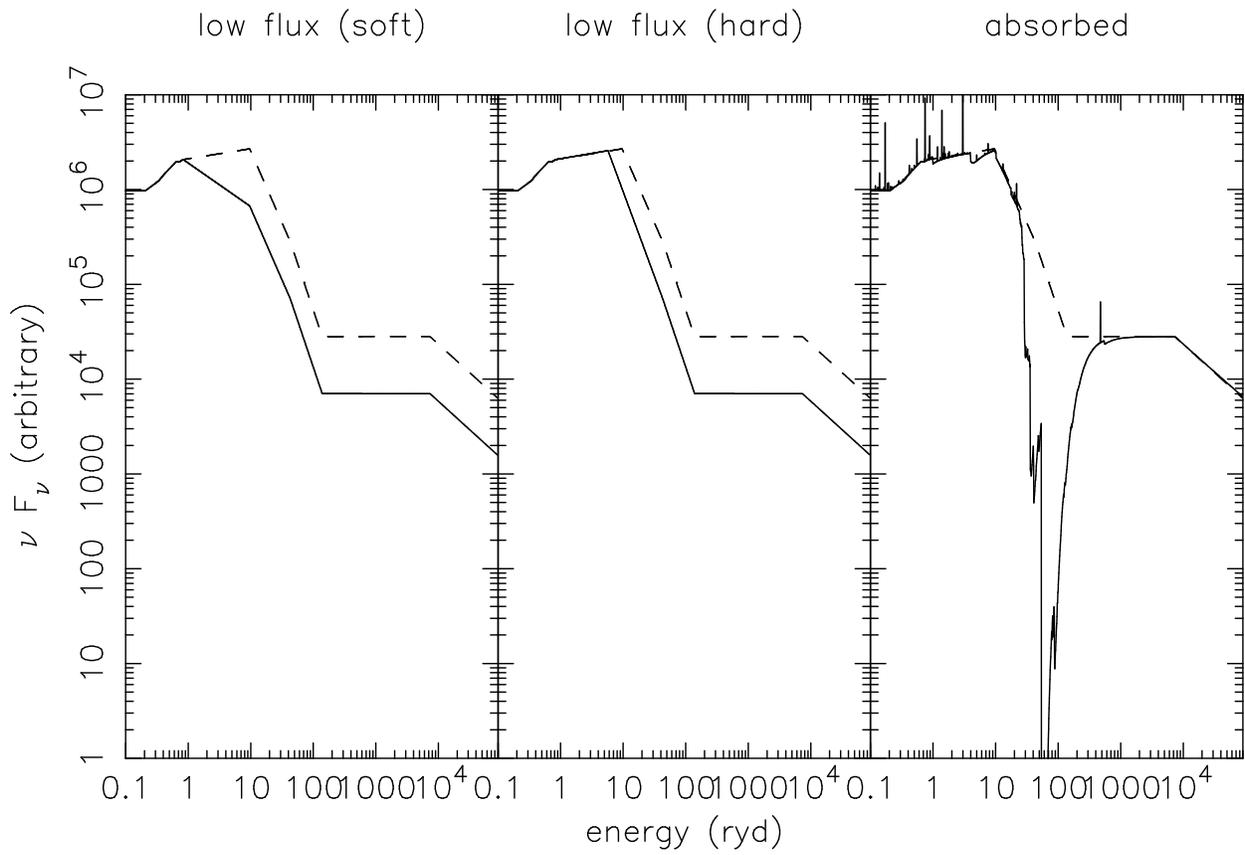}
\caption{Portions of the continua used for modeling the wind, compared
in each case with the IRAS~13224$-$3809 continuum developed from the
{\it HST} and non-simultaneous ASCA and ROSAT spectra.\label{fig2}}
\end{figure}

\clearpage

\begin{figure}
\epsscale{0.8}
\plotone{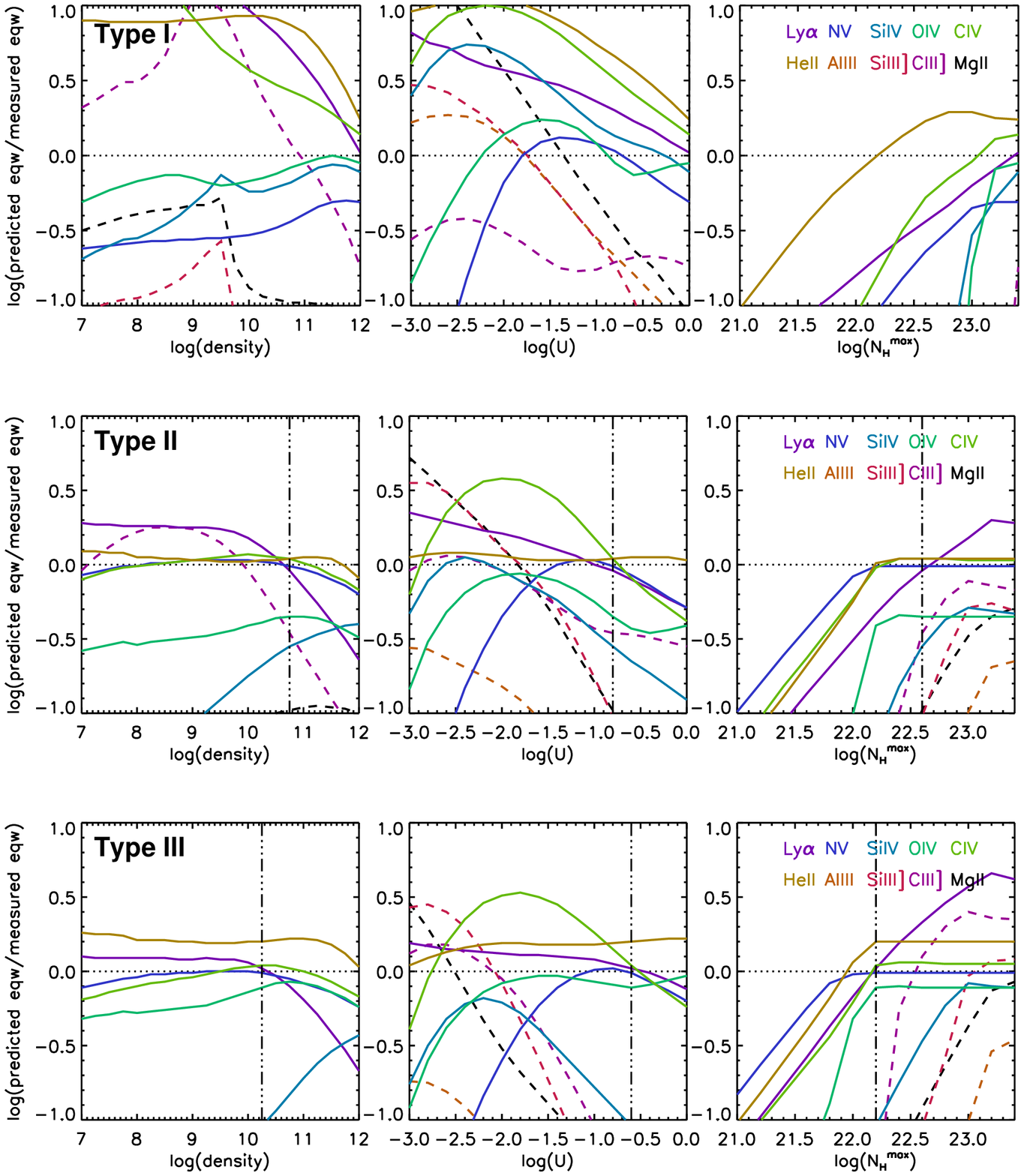}
\caption{Examples of the three representative types of solutions for
  the photoionization modeling of the wind.  The log of the ratio
  of the observed equivalent width to the   predicted   equivalent
  width of Ly$\alpha$, \ion{N}{5}, \ion{Si}{4},   \ion{O}{4},
  \ion{C}{4}, and \ion{He}{2} (solid lines), and the log   of the
  ratio of the upper limits of the equivalent widths to the
  predicted equivalents of \ion{Al}{3}, \ion{Si}{3}], \ion{C}{3}], and
  \ion{Mg}{2} (dashed lines) is plotted.  The dashed-dotted line marks
  the position of the solution, as follows.  The Type I representative
  solution consists of the nominal continuum, solar metallicity,
  $\log(n)=12$, $\log(U)=0$, $\log(N_H^{max})=23.4$, and a covering
  fraction of 0.26.  The Type II representative solution consists of
  the nominal continuum, and metals$\times 5$ and 
  nitrogen$\times 10$ metallicity, and $\log(n)=10.75$, $\log(U)=-0.8$,
  $\log(N_H^{max})=22.6$, and covering fraction of 0.08.  The Type III
  representative solution  has the hard low-flux continuum,
  and metals$\times 5$ and nitrogen$\times 10$ metallicity, and
  $\log(n)=10.25$, $\log(U)=-0.6$, $\log(N_H^{max})=22.2$, and
  covering fraction of 0.15.\label{fig3}}
\end{figure}

\clearpage

\begin{figure}
\epsscale{1.0}
\plotone{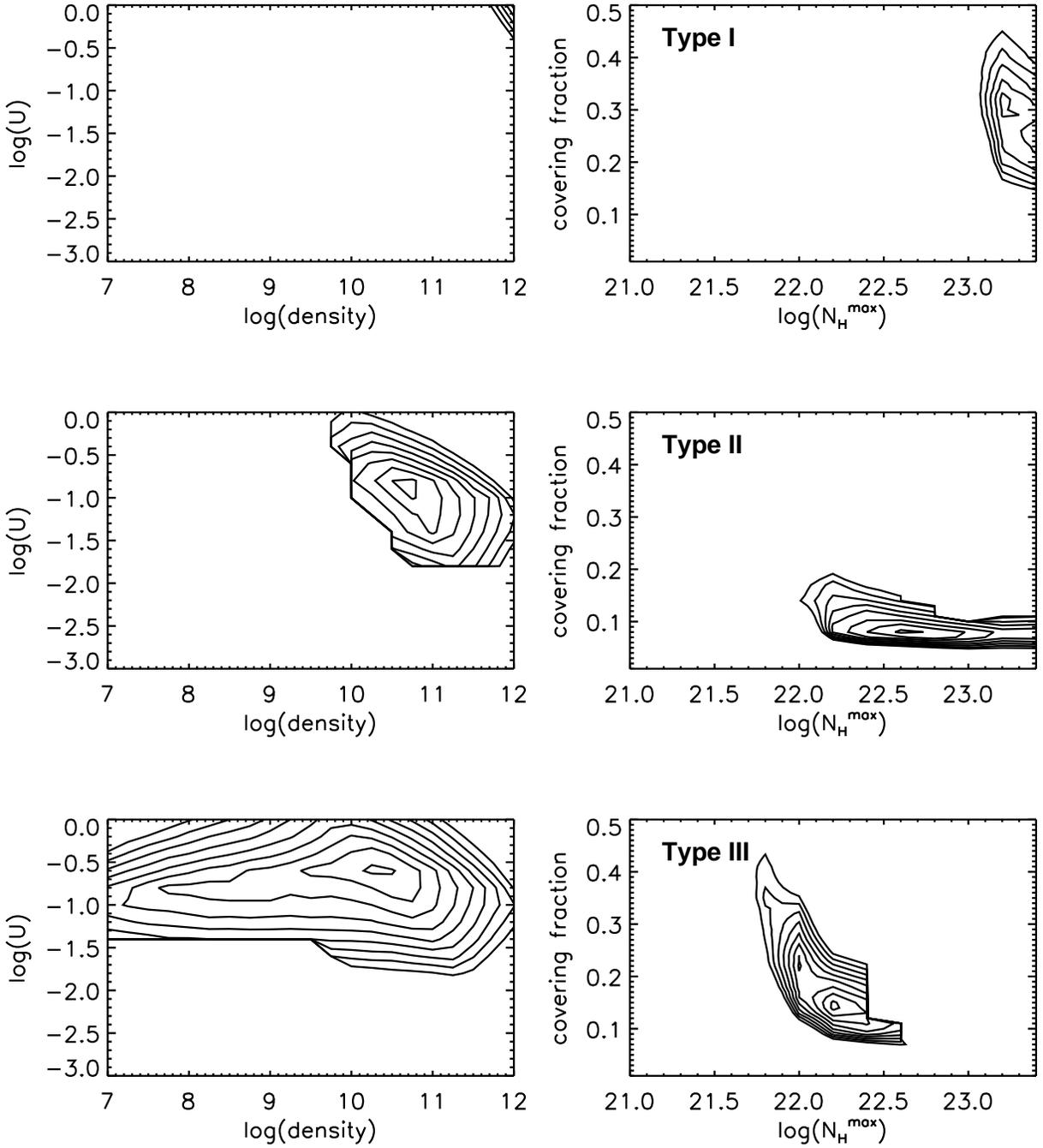}
\caption{Contours of {\it fom} for the three examples of the three
  representative types of solutions for 
  the photoionization modeling of the wind described in the text and
  in Fig.\ 3. In each case, the contour levels are {\it fom}=0 to 120
  with interval 10.  The
  continuum, metallicity and minimum {\it fom} parameter values are
  listed in the text and the caption of Fig.\ 3.\label{fig4}}
\end{figure}

\clearpage

\begin{figure}
\epsscale{0.7}
\plotone{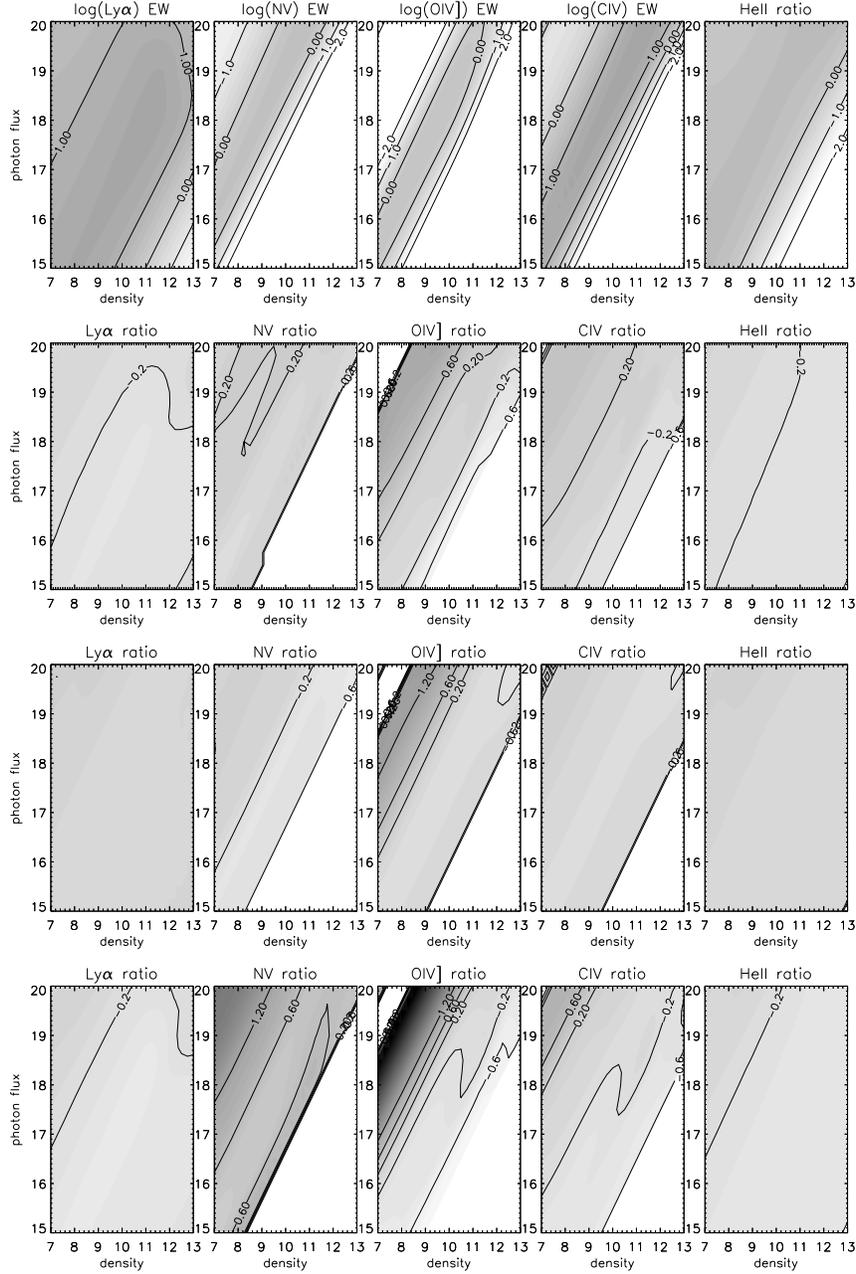}
\caption{Top: Equivalent widths for selected lines for a model
  consisting of the nominal continuum, solar metalicity,
  $\log(N_H^{max})=22.4$, and a covering fraction of 0.1.  Second: Log of
  the ratio of emission lines for a model having  the metallicity
  enhanced by a factor of 5 to those from the model with nominal
  continuum and solar metallicity.  $N_H^{max}$ is adjusted so that
  the fraction of the  distance to the hydrogen ionization front is
  the same in both cases.  Third: Log of the ratio of the emission
  lines for a model having the hard low-flux continuum with solar
  metallicity to those from the nominal continuum and solar
  metallicity. Bottom: Combines results from second and third panels,
  showing the log of the ratio of the emission lines for a model
  having the hard low-flux continuum and with metals enhanced by a
  factor of 5 times solar, and nitrogen enhanced by a factor of 10
  times solar to those from a model with nominal continuum and nominal
  metallicity.\label{fig5}}
\end{figure}

\clearpage

\begin{figure}
\epsscale{0.5}
\plotone{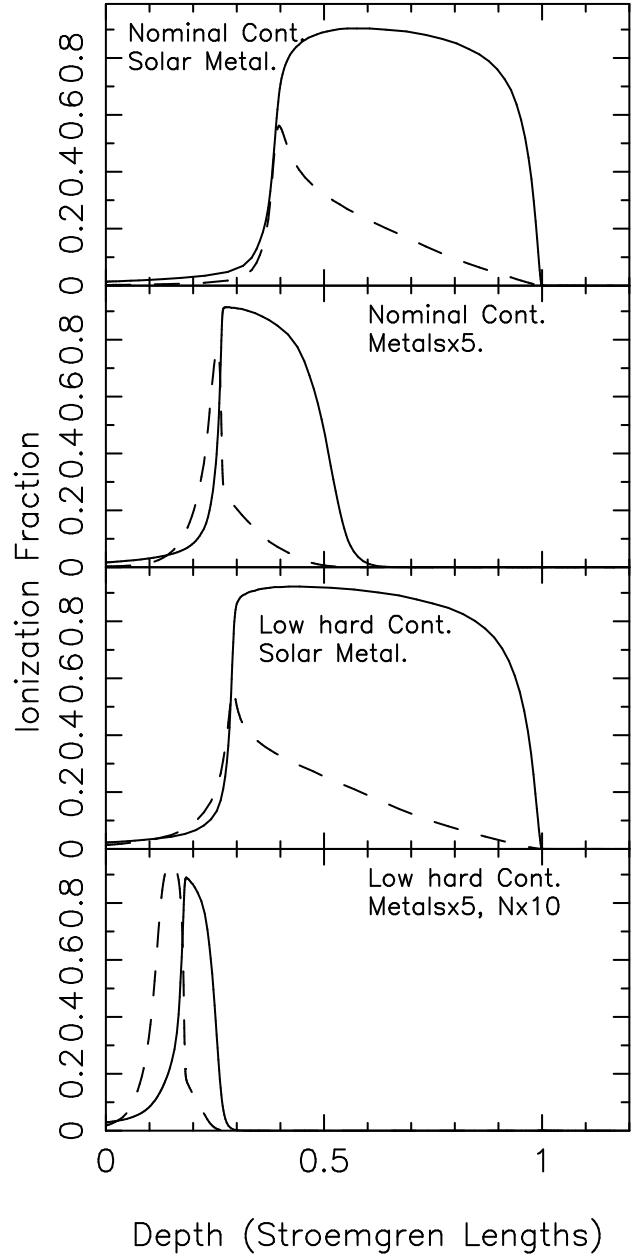}
\caption{Ionization fractions for C+3 (solid line) and O+3 (dashed
  line) for different continua and metallicity as a function of depth
  into the ionized gas normalized by the depth to the hydrogen
  ionization front.  \label{fig6}}
\end{figure}

\clearpage

\begin{figure}
\epsscale{1.0}
\plotone{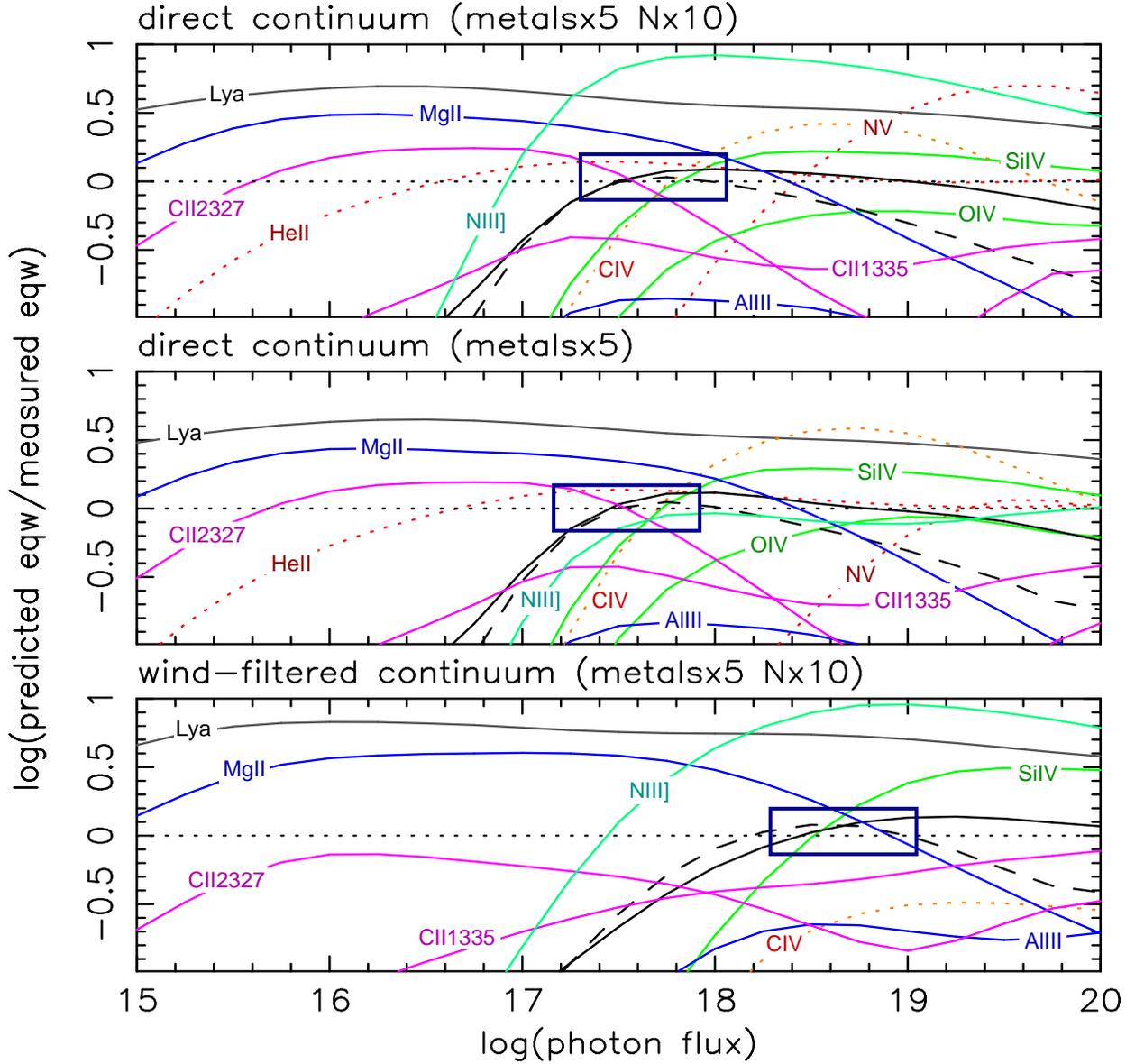}
\caption{Results of photoionization modeling for disk. The log of the
ratio of the model line equivalent widths with the measured equivalent
widths are shown.  The density and covering fraction were first determined
using the equivalent widths of \ion{C}{3}] (solid black line) and
\ion{Si}{3}] (dashed black line) and their ratio, as described in the
text.  Other lines are marked: solid lines denote detection, and
dotted lines denote upper limits.  Boxes mark the approximate range of
solutions approximately consistent with the observed emission
lines.\label{fig7}}
\end{figure}

\clearpage

\begin{figure}
\epsscale{1.0}
\plotone{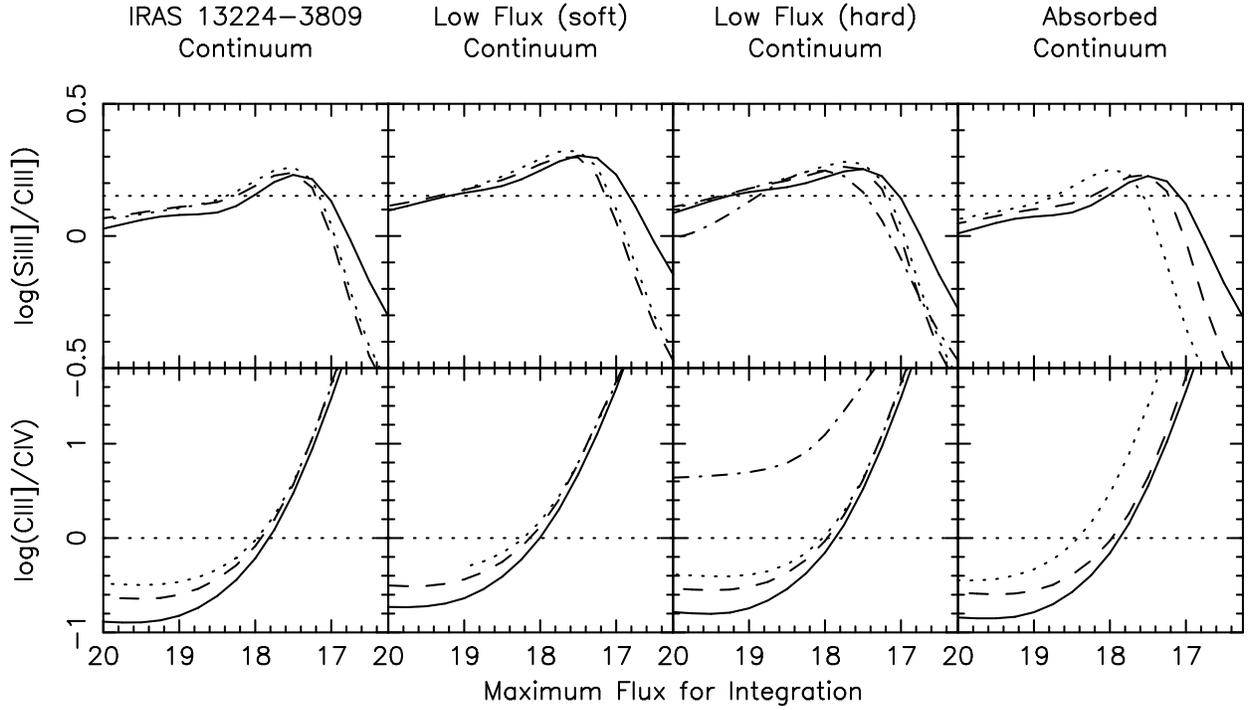}
\caption{The logarithm of the ratios of \ion{Si}{3}] to \ion{C}{3}]
and \ion{C}{3}] to \ion{C}{4} for different continua (panels left to
right) and metallicities (solid line -- solar; dashed line --
metals$\times5$; dotted line -- metals$\times5$ and N$\times 10$).
The dot-dashed line shows the wind-filtered continuum.  In
each case, the line emission is integrated from the radius
corresponding to the flux on the x-axis (the minimum radius) to
$\log(\Phi)=15$ (the maximum radius).  See text for additional
information.  Horizontal dotted lines show the measured value, for the
\ion{Si}{3}]/\ion{C}{3}] ratio, and the lower limit, for the
\ion{C}{3}]/\ion{C}{4} ratio.\label{fig8}}
\end{figure}

\clearpage

\begin{figure}
\epsscale{1.0}
\plotone{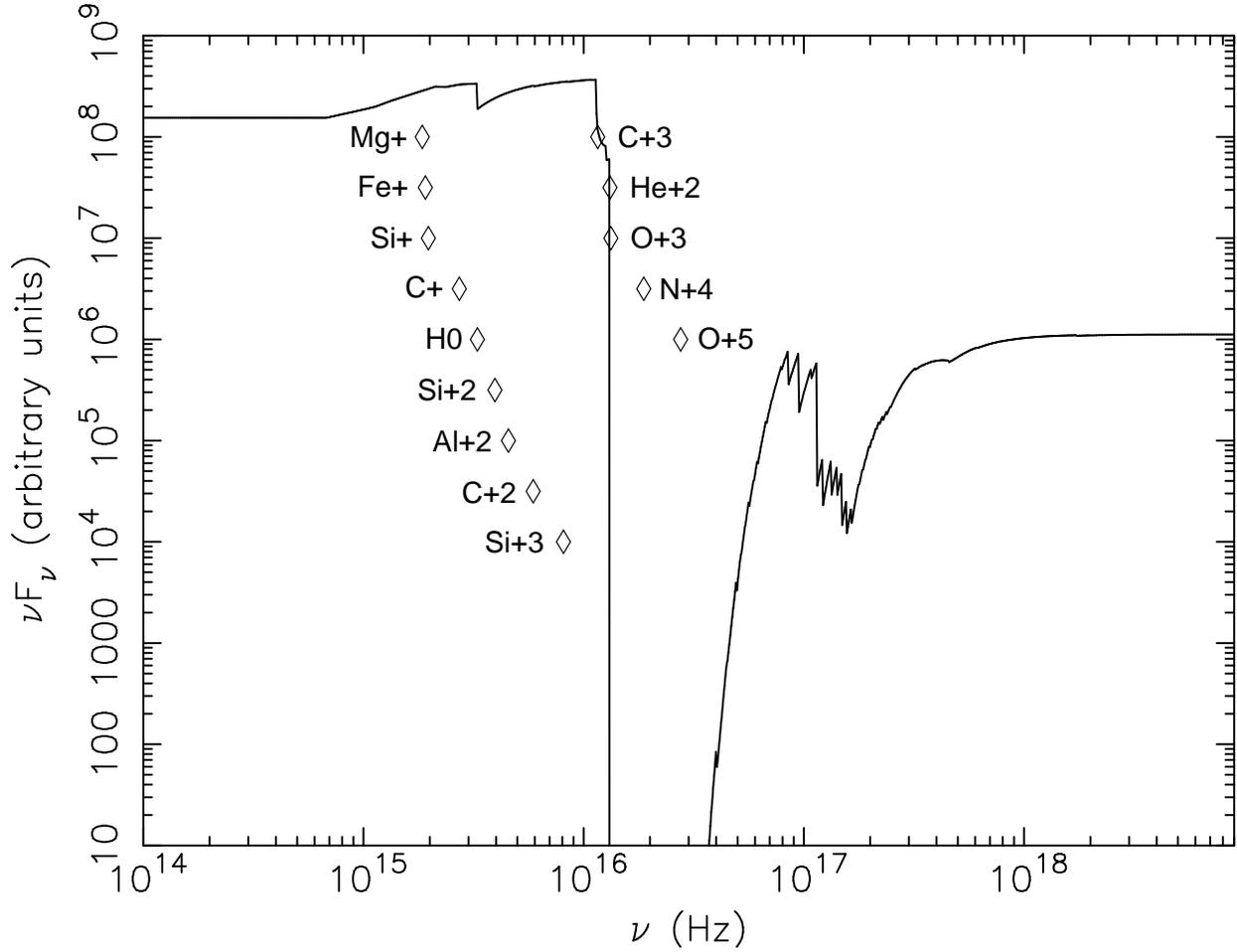}
\caption{The ``filtered'' continuum, produced as a result of
transmitting the hard low-flux continuum through the wind. The
frequencies corresponding to the ionization potentials of various
ions are marked.   The lack of emission in the helium continuum is
seen.\label{fig9}} 
\end{figure}

\clearpage

\begin{figure}
\epsscale{1.0}
\plotone{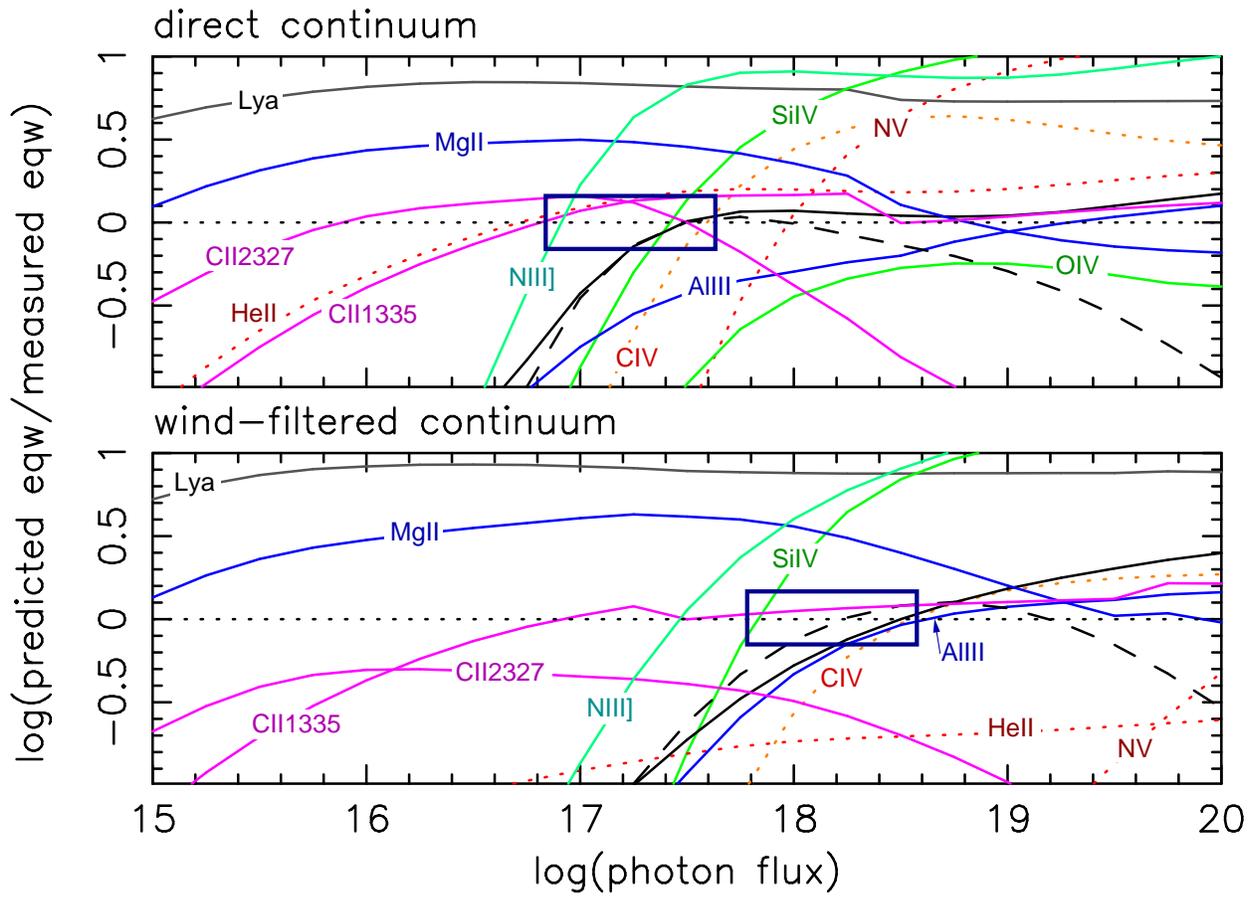}
\caption{Same as top and bottom panels of Fig.\ 7, with turbulence of
  $2,000\rm\, km\,s^{-1}$.\label{fig10}} 
\end{figure}

\clearpage

\begin{figure}
\epsscale{1.0}
\plotone{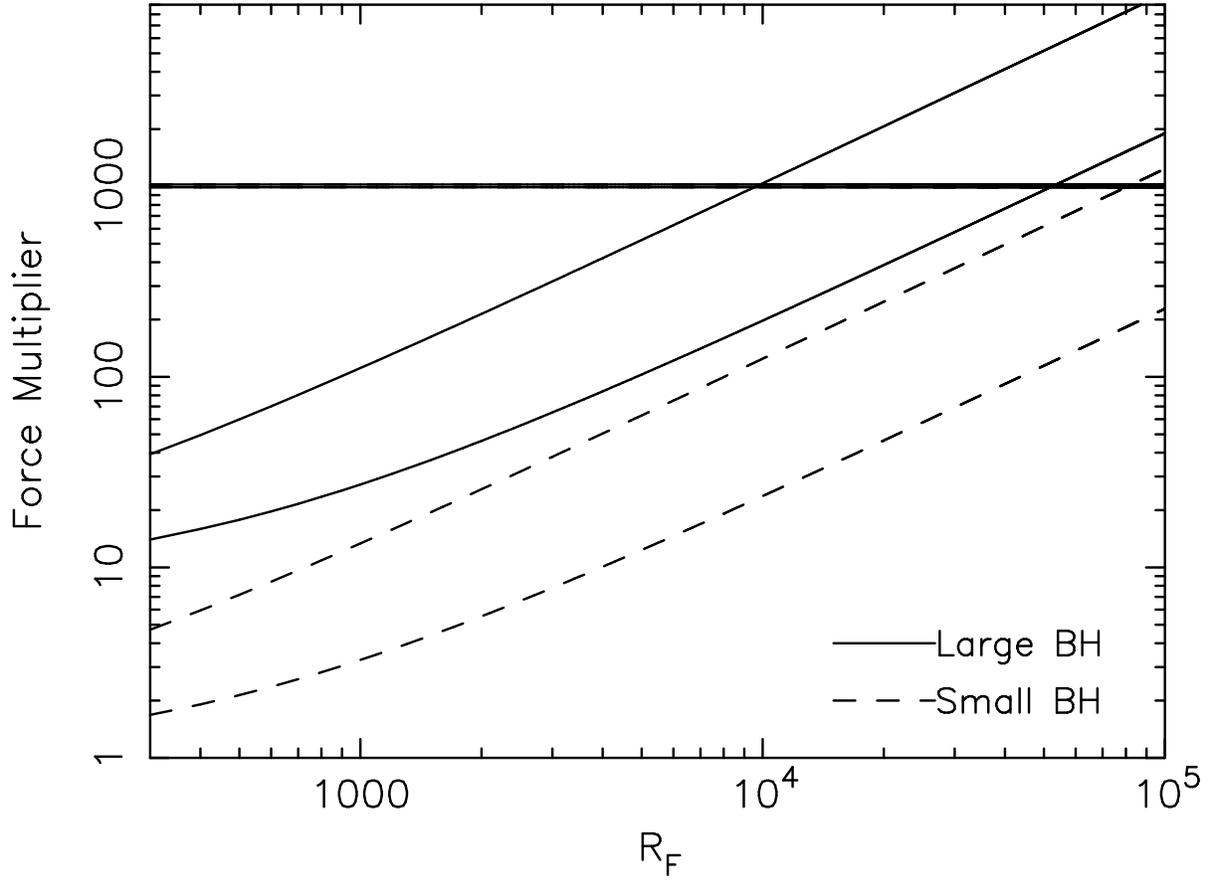}
\caption{Illustration of kinematic solutions for the wind.  Diagonal
  lines are the force multiplier required to attain a terminal
  velocity projected onto the symmetry axis of $10,000\rm\, 
  km\,s^{-1}$. Solid and dashed lines correspond to the large and small black
  hole masses, respectively.  For each assumed black hole mass, the
  upper and lower lines correspond to cosine of the stream line angle
  with respect to the symmetry axis of 0.3 and 0.7, respectively.  The
  horizontal lines are the values of the force multiplier implied by
  the physical state of the gas.  The point where the two force
  multiplier estimates  cross is a self-consistent
  solution. \label{fig11}} 
\end{figure}

\clearpage

\begin{figure}
\epsscale{0.8}
\plotone{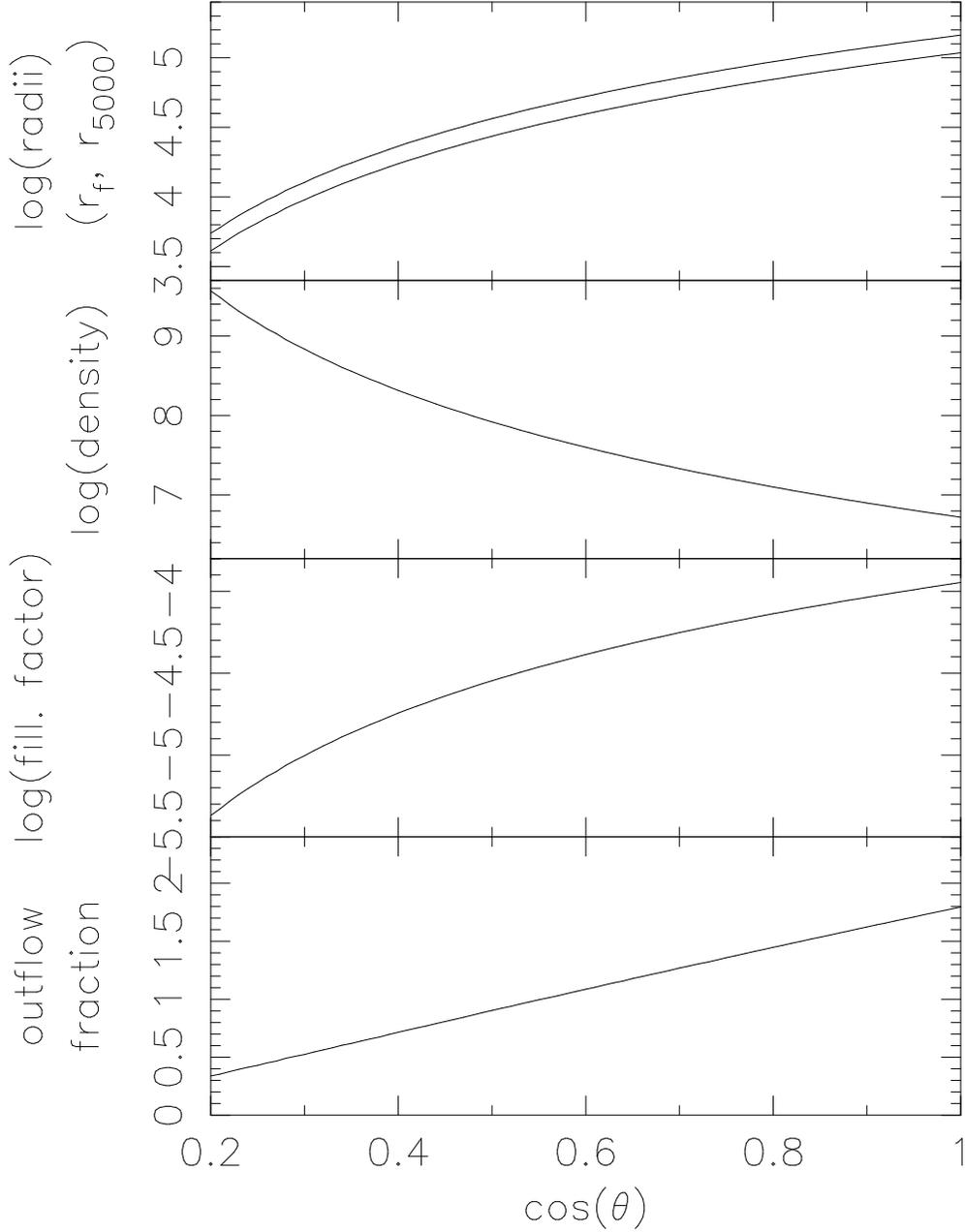}
\caption{Model results for $L/L_{edd}=0.12$ and the hard low-flux continuum as
a function of $\cos(\Theta)$, where $\Theta$ is the angle between the
symmetry axis and the wind streamlines.  Top: Foot-point radius and radius
at which the projected wind velocity reaches $5,000\rm\, km\,s^{-1}$,
as a function of Schwarzschild radius for a $1.3 \times 10^{8}\rm
M_\odot$ black hole; Middle-top: density of the outflow;
Middle-bottom: filling factor for the outflow; Bottom:
\.M$_{wind}$/\.M$_{acc}$, assuming the covering
fraction of 0.15 obtained from photoionization modeling, and assuming
that the efficiency of conversion of gravitational potential energy to
radiation is 10\%.\label{fig12}}
\end{figure}

\clearpage

\begin{figure}
\epsscale{1.0}
\plotone{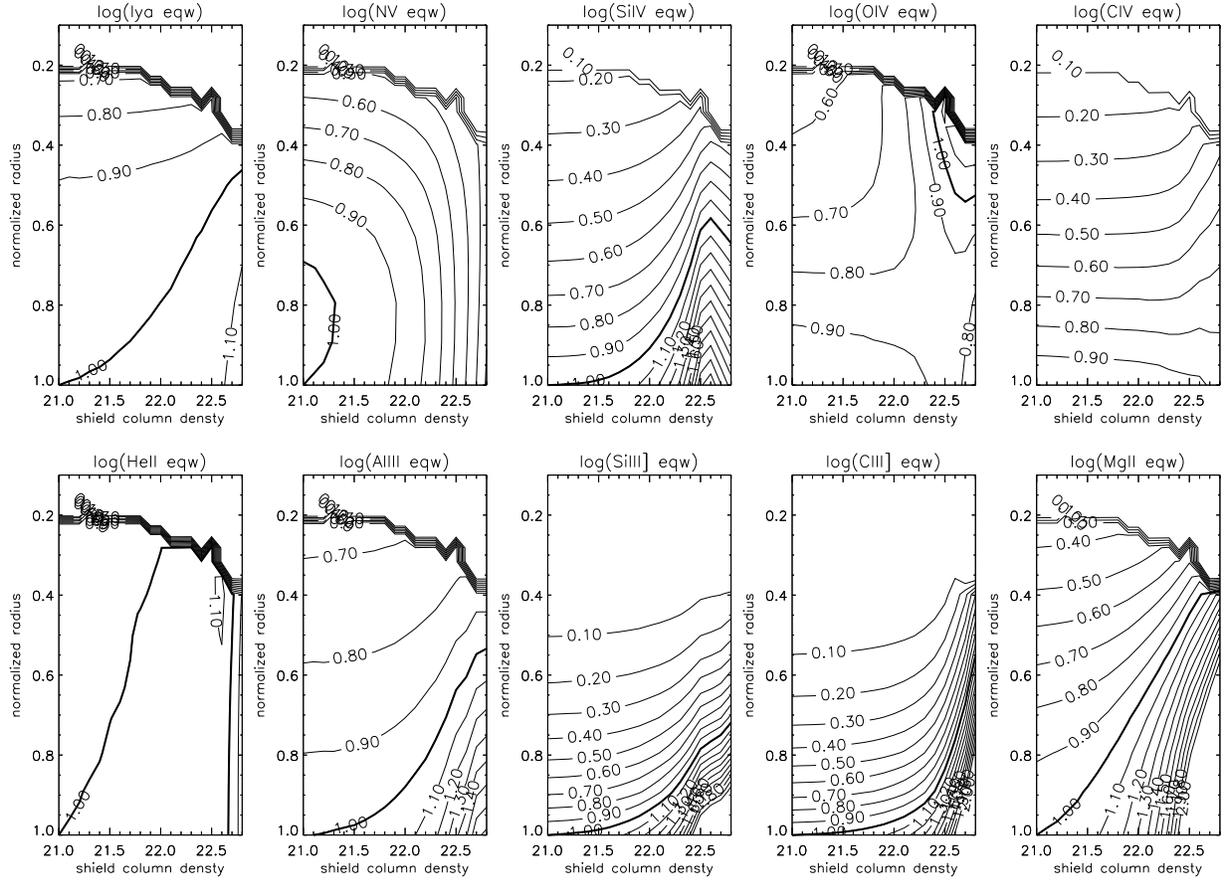}
\caption{Test of the effect of ``shielding''.  Contours of normalized
  equivalent width, constructed with a covering fraction of 0.2.
  Normalization has been done with respect to the nominal solution at
  the lower left corner of each plot; this is nearly equivalent to the
  solution obtained in \S 2.2.  The x-axis shows the column density of
  the shielding gas, which has an ionization parameter of U=10.  The
  y-axis plots the radius with respect to the nominal value in the
  lower left corner of each plot.  The thick contour intersecting the
  lower left corner of each plot shows the nominal value for each
  emission line.   See text for further details.\label{fig13}} 
\end{figure}

\clearpage
\begin{figure}
\epsscale{0.5}
\plotone{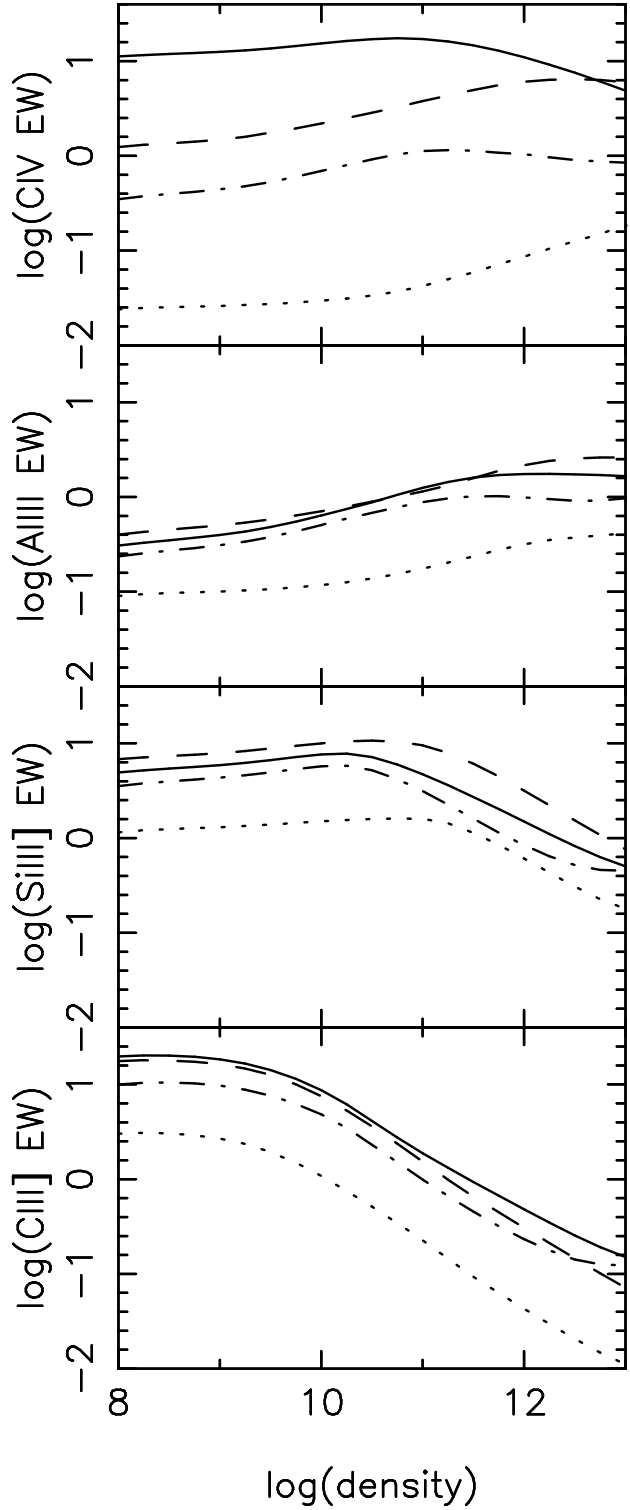}
\caption{Illustration of mutual influence of density, spectral shape,
and ionization parameter for \ion{C}{4}, \ion{Al}{3},
\ion{Si}{3}], and \ion{C}{3}].  The hard 
  low-flux continuum was assumed, and the assumed covering fraction is
0.05.  For each emission line, the solid line denotes $\log(U)=-2$
and direct illumination, the dashed line denotes $\log(U)=-3$ and
direct illumination, the dash-dotted line denotes $\log(U)=-2$ and a
filtered continuum,  and the dotted line denotes
$\log(U)=-3$ and a filtered continuum. \label{fig14}}
\end{figure}

\clearpage

\begin{figure}
\epsscale{1.0}
\plotone{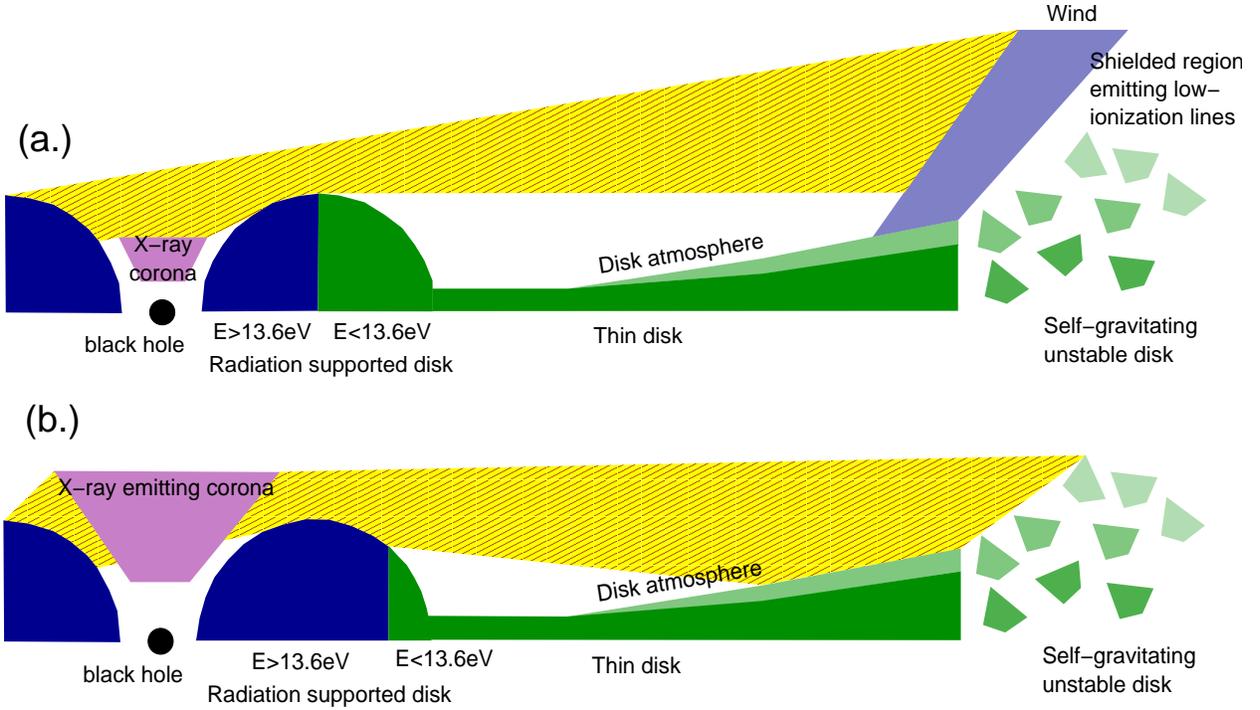}
\caption{Schematic diagram for a speculative scenario referenced in \S
  4.5.1 (not to scale). Red diagonal 
lines with yellow background show rays of photoionizing light.  {\it
(a.)} This geometry encourages a wind to develop, which then shields
the intermediate- and low-ionization region possibly associated with
the self-gravitating, unstable disk (Collin \& Hur\'e 2001).  The
outside of the radiation supported disk emits primarily light that
initially accelerated but does not overionize the wind. {\it (b.)}
This geometry does not produce a wind.  A significant amount of
photoionizing light is emitted by the outside of the
radiation-supported disk.  This overionizes the tenuous part of the
disk atmosphere, preventing a wind from being formed. The disk
atmosphere is illuminated by photoionizing light, causing
high-ionization lines to be formed that will be symmetric and centered
at the rest wavelength.\label{fig15}}
\end{figure}

\clearpage

\begin{deluxetable}{llllllll}
\tabletypesize{\scriptsize}
\tablewidth{0pt}
\tablecaption{Wind Solution Results}
\tablehead{
\colhead{Continuum} & \colhead{Metallicity$^a$} & \colhead{Minimum {\it
    fom}} & \colhead{\# {\it fom}$<120$} & \colhead{$\log(n)^b$} &
\colhead{$\log(U)^b$} & \colhead{$\log(N_H^{max})^b$} &
\colhead{Covering Fraction$^b$}}
\startdata
nominal & solar & 75 & 211 & 11.75 -- 12.0 & $-0.6$ -- 0.0 & 23.0 --
23.4 & 0.14 -- 0.4 \\
nominal & metals$\times 5$ & 57 & 2612 & 10.0 -- 12.0 & $-1.4$ -- 0.0 &
22.0 -- 23.4 & 0.07 -- 0.41 \\
nominal & metals$\times 5$ N$\times 10$ & 11 & 12049 & 7.0 -- 12.0 &
$-2.0$ -- 0.0 & 21.4 -- 23.4 & 0.06 -- 0.48 \\
low flux soft & solar & 74 & 756 & 11.5 -- 12.0 & $-0.8$ -- 0.0 &
22.8 -- 23.4 & 0.2 -- 0.49 \\
low flux soft & metals$\times 5$ & 59 & 4609 & 8.5 -- 12.0 & $-1.2$ --
0.0 & 22.0 -- 23.4 & 0.18 -- 0.49 \\
low flux soft & metals$\times 5$ N$\times 10$ & 16 & 17817 & 7.0 --
12.0 & $-1.6$ -- 0.0 & 21.6 -- 23.4 & 0.16 -- 0.49 \\
low flux hard & solar & 66 & 600 & 11.5 -- 12.0 & $-0.6$ -- 0.0 &
22.8 -- 23.4 & 0.14 -- 0.47 \\
low flux hard & metals$\times 5$ & 61 & 5722 & 7.75 -- 12.0 & $-1.2$ --
0.0 & 21.6 -- 23.4 & 0.09 -- 0.49 \\
low flux hard & metals$\times 5$ N$\times 10$ & 28 & 18400 & 7.0 --
12.0 &  $-1.8$ -- 0.0 & 21.4 -- 23.4 & 0.09 -- 0.49 \\
ionized absorber$^c$ & solar & 76 & 226 & 11.5 -- 12.0 & $-0.6$ -- 0.0 &
23.0 -- 23.4 & 0.13 -- 0.5\\
ionized absorber$^c$ & metals$\times 5$ & 61 & 2176 & 10.75 -- 12.0 & $-1.4$
-- 0.0 & 22.2 -- 23.4 & 0.06 -- 0.5 \\
ionized absorber$^c$ & metals$\times 5$ N$\times 10$ & 12 & 12454 & 7.0 --
12.0 & $-2.2$ -- 0.0 & 21.2 -- 23.4 & 0.05 -- 0.5 \\
\enddata
\tablenotetext{a}{Metallicity value of ``metals$\times 5$'' means that
  the metals are assumed enhanced by a factor of 5 over solar.
  ``metals$\times 5$ N$\times 10$''  means that metals are assumed enhanced
  by a factor of 5 over solar, and nitrogen alone is enhanced by a
  factor of 10 over solar.}
\tablenotetext{b}{The range of parameters listed represent 95\% of the
full range of parameter combinations with {\it fom}$<120$.}
\tablenotetext{c}{The ionization parameter listed is the value before
  transmission through the absorber.}
\end{deluxetable}

\clearpage

\begin{deluxetable}{lllll}
\tabletypesize{\scriptsize}
\tablewidth{0pt}
\tablecaption{Disk Photoionization Model Parameters}
\tablehead{
\colhead{Continuum} & \colhead{$\log($density)} & \colhead{$\log($Photon
Flux$)^a$} & \colhead{$\log(U)^a$} & \colhead{Covering Fraction} \\}
\startdata
direct (Metals$\times5$ N$\times10$) & 10.0 &  17.5 & $-3.0$ & 0.035 \\
direct (Metals$\times5$) & 10.0 &  17.5 & $-3.0$ & 0.03 \\
wind-filtered (Metals$\times5$ N$\times10$) & 10.25 & 18.7 & $-2.1$ & 0.05 \\
direct (Metals$\times5$ N$\times10$) $v_{turb}=2000\rm\,km\,s^{-1}$ & 10.0 &  17.25 & $-3.2$ & 0.035 \\
wind-filtered (Metals$\times5$ N$\times10$) $v_{turb}=2000\rm\,km\,s^{-1}$  & 10.25 & 18.2 & $-2.5$ & 0.05 \\
\enddata
\tablenotetext{a}{The  photon flux and ionization parameter listed for
filtered continuum are the values appropriate for the flux {\it
  before} it is transmitted through the absorber.}

\end{deluxetable}



\end{document}